\documentclass[pre,twocolumn,nobalancelastpage,amsmath,amssymb,showpacs,superscriptaddress,reprint,
nofootinbib]{revtex4-1}

\usepackage{graphics}      
\usepackage{graphicx}      
\usepackage{longtable}     
\usepackage{url}           
\usepackage{bm}            
\usepackage{xcolor}
\usepackage{amsmath}
\usepackage{dcolumn}

\newcommand{\Eq}[1]{Eq.~(\ref{eq:#1})}
\newcommand{\rr}{\mathbf{r}}
\newcommand{\B}{\mathbf{B}}
\newcommand{\Y}{\mathbf{Y}}
\newcommand{\rji}{\rr_{ji}}
\newcommand{\bnabla}{\boldsymbol{\nabla}}

\usepackage{pgf, tikz}
\usetikzlibrary{shapes.misc}
\usetikzlibrary{decorations.pathreplacing}

\begin{document}
\title{Orientational Ordering in Athermally Sheared, Aspherical, Frictionless Particles}
\author{Theodore Marschall}
\affiliation{Department of Physics and Astronomy, University of Rochester, Rochester, NY 14627}
\author{Yann-Edwin Keta}
\affiliation{Department of Physics, Ume{\aa} University, 901 87 Ume{\aa}, Sweden}
\affiliation{D\'epartement de Physique, \'Ecole Normale Sup\'erieure de Lyon, 69364 Lyon Cedex 07, France}
\affiliation{D\'epartement de Physique, Universit\'e Claude Bernard Lyon 1, 69622 Villeurbanne Cedex, France}
\author{Peter Olsson}
\affiliation{Department of Physics, Ume{\aa} University, 901 87 Ume{\aa}, Sweden}
\author{S. Teitel}
\affiliation{Department of Physics and Astronomy, University of Rochester, Rochester, NY 14627}
\date{\today}

\begin{abstract}
We   numerically simulate the uniform athermal  shearing  of bidisperse, frictionless, two dimensional  spherocylinders and three dimensional  prolate ellipsoids. We focus on the orientational ordering of  particles
as an asphericity parameter $\alpha\to 0$ and particles approach spherical.  We find that  the nematic order parameter $S_2$ is  non-monotonic  in the packing fraction $\phi$, and that as
$\alpha\to 0$ $S_{2}$  stays finite at jamming and above. 
The approach to spherical particles thus appears to be singular.  We also find that  sheared particles continue to rotate above jamming, and that particle contacts  preferentially lie along the narrowest width of the particles, even as $\alpha\to 0$. 
\end{abstract}
\maketitle


Models of athermal ($T=0$) granular materials have often focused on the simplest case of spherical particles.  Recently, however, more attention has  been paid to the case of   elongated particles with lower rotational symmetry \cite{Borzsonyi.Soft.2013}.  The question then  arises whether   such elongated particles will  orientationally order as the particle density increases, so as to pack more efficiently.  While elongated particles in thermal equilibrium are known to have a nematic orientational ordering transition \cite{Onsager,Bolhuis},  recent works have found that such particles do {\em not} orientationally order upon athermal  isotropic compression \cite{Donev.Science.2004,Man.PRL.2005,Sacanna.JPhysC.2007,Zhao,Marschall}.

Orientational ordering is, however,  found when  elongated particles are placed in an athermal uniform shear flow.  In this case,
drag forces between the particle and the flow will  cause the particle to tumble \cite{Jeffery}.  If the particle is asymmetrical, with unequal eigenvalues of its moment of inertia tensor,  tumbling will have a non-uniform rotational velocity; the torque from drag forces will vary with the particle's orientation, and the particle will on average  align with the flow direction.  For a finite density of colliding particles,   nematic ordering remains but the nematic director becomes oriented at a finite angle with respect to the flow direction \cite{Campbell,Guo1,Guo2,Borzsonyi1,Borzsonyi2,Wegner,Wegner2,Nagy,Trulsson}.

Here we investigate the  nematic ordering of frictionless,  aspherically shaped  particles,  athermally sheared at  constant  strain rate $\dot\gamma$, putting the system into a steady state of simple shear flow.
We consider behavior as an asphericity parameter $\alpha\to 0$, and the particles approach spherical. 
We find the surprising result that a {\em finite} nematic ordering persists even as $\alpha\to 0$,
suggesting that the shear driven jamming of aspherical particles has a singular limit as $\alpha\to 0$.
Since most  particles in nature are not truly spherical, our result may have broad implications for granular shear flows.

{\bf Models:} We consider two different numerical models: (i) spherocylinders in two dimensions (2D), and (ii) prolate ellipsoids in three dimensions (3D).  In both cases we take a bidisperse distribution of particle sizes, with equal numbers of big and small particles.  We use soft-core particles with a one-sided harmonic elastic repulsion.  The system  length is $\mathcal{L}$ in all directions, with periodic boundary conditions along the flow direction  $\mathbf{\hat x}$, and  Lees-Edwards boundary conditions \cite{LeesEdwards} with a uniform  strain rate $\dot\gamma$ in the transverse direction $\mathbf{\hat y}$.  In 3D  we take periodic boundary conditions along $\mathbf{\hat z}$.  The particle packing fraction is $\phi=\sum_i v_i/\mathcal{V}$, with $v_i$  the volume of particle $i$ and $\mathcal{V}=\mathcal{L}^d$  the system volume ($d=2$ or $3$ for 2D and 3D respectively).

 {\bf 2D Spherocylinders:} A 2D spherocylinder  consists of a rectangle of length $L$, with two semi-circular end caps of diameter $D$ (see inset to Fig.~\ref{f5}a).  We define the asphericity parameter  $\alpha=L/D$. Big and small particles have equal  $\alpha$, with $D_b/D_s=1.4$.
Taking the ``spine" of the spherocylinder as the line  bisecting the rectangle parallel to its length $L$,   we define $r_{ij}$ as the shortest distance between the spines of spherocylinders $i$ and $j$ and $d_{ij}=(D_i+D_j)/2$.  Two spherocylinders are  in contact whenever $r_{ij}<d_{ij}$, in which case the  elastic interaction is $U^\mathrm{el}=(k_e/2)(1-r_{ij}/d_{ij})^2$ and the  repulsive elastic force on $i$ is  
$\mathbf{F}^\mathrm{el}_{ij}=(k_e/d_{ij})(1-r_{ij}/d_{ij})\mathbf{\hat n}_{ij}$, with $\mathbf{\hat n}_{ij}$ the unit vector pointing normally inwards to particle $i$ at the point of contact with  $j$ \cite{Marschall,SupMat}.  

Our dynamics  is   the mean-field Durian  model for foams \cite{Durian}, generalized to non-spherical particles. 
The dissipative force on a spherocylinder is a Stokes drag between the particle and a uniform background shear flow, $\mathbf{F}_i^\mathrm{dis}=-k_d v_i(\dot{\mathbf{r}}_i-y_i\dot\gamma\mathbf{\hat x})$, with $\mathbf{r}_i=(x_i,y_i)$  the center of mass of spherocylinder $i$, $\dot{\mathbf{r}}_i$ the center of mass velocity, and $k_d$ the viscous coupling.  We use  overdamped dynamics $\mathbf{F}_i^\mathrm{dis}+\sum_j\mathbf{F}_{ij}^\mathrm{el}=0$, where the sum is over all particles $j$ in contact with $i$.

The elastic and dissipative forces produce torques on the spherocylinders.
 The elastic torque  on particle $i$ due to contact with  $j$  is, $\boldsymbol{\tau}_{ij}^\mathrm{el}=\mathbf{\hat z}\tau_{ij}^\mathrm{el}=\mathbf{s}_{ij}\times \mathbf{F}_{ij}^\mathrm{el}$, where $\mathbf{s}_{ij}$ is the moment arm from the center of mass of  $i$ to its point of contact with $j$.  A dissipative torque arises from  the variation of the background shear flow velocity over the spatial extent of the particle \cite{Marschall2}.  Integrating over  particle area gives $\tau_i^\mathrm{dis}=-k_dv_i I_i[\dot\theta_i+\dot\gamma f(\theta_i)]$, where $\theta_i$ is the angle of the  spine with respect to the flow direction $\mathbf{\hat x}$, and $f(\theta)=[1-C\cos2\theta]/2$.  The overdamped $\tau_i^\mathrm{dis}+\sum_j \tau_{ij}^\mathrm{el}=0$  determines the particle rotation.  Here $I_i$ is the sum of the two eigenvalues of the moment of inertia tensor, and $C=\Delta I_i/I_i$, with $\Delta I_i$  the difference between the two eigenvalues.  For spherocylinders, $I_i=(D_i/2)^2(3\pi+24\alpha+6\pi\alpha^2+8\alpha^3)/(6\pi+24\alpha)$.
For a circle, $\Delta I=0$, and so in the absence of collisions  $\dot\theta/\dot\gamma=-1/2$.  We take  as unit of length $D_s=1$,  unit of energy $k_e=1$, and  unit of time $t_0=D_s^2k_d/k_e=1$.  We integrate the equations of motion using the Heun method with  step size $\Delta t/t_0=0.02$.  We use $N=1024$ particles.  

{\bf 3D Prolate Ellipsoids:} We take prolate ellipsoids of revolution with major axis  length $a_1$ and minor axes  length $a_2$.   The asphericity  is $\alpha=a_1/a_2-1$.  Big and small particles have equal $\alpha$, with $a_{1b}/a_{1s}=1.4$. When two ellipsoids $i$ and $j$  overlap, we define a scale factor $\delta_{ij}<1$ such that
the particles  just barely touch when their axes are rescaled by $\delta_{ij}$, keeping the center of mass positions fixed \cite{SupMat}. 
The elastic interaction is then 
$U^\mathrm{el}=(k_e/2)(1-\delta_{ij})^2$, 
and the  repulsive elastic force on $i$ is $\mathbf{F}_{ij}^\mathrm{el}=k_e \delta_{ij} (1-\delta_{ij})\mathbf{\hat n}_{ij}/[(\mathbf{r}_i-\mathbf{r}_j)\cdot\mathbf{\hat n}_{ij}]$, with $\mathbf{r}_i$  the center of mass of ellipsoid $i$ and $\mathbf{\hat n}_{ij}$ the unit vector pointing normally inwards to particle $i$ at the point of contact with  $j$.

We take a purely collisional dynamics.  The dissipative force on ellipsoid $i$ is due to contact with  $j$ and  is proportional to the difference in particle velocities at their point of contact, $\mathbf{F}_{ij}^\mathrm{dis}=-k_d (\dot{\mathbf{r}}_i+\boldsymbol{\omega}_i\times \mathbf{s}_{ij}-\dot{\mathbf{r}}_j-\boldsymbol{\omega}_j\times\mathbf{s}_{ji})$, with $\dot{\mathbf{r}}_i$  the center of mass velocity, $\boldsymbol{\omega}_i$  the angular velocity about the center of mass, and $\mathbf{s}_{ij}$  the moment arm from the center of  $i$ to the point of contact with  $j$ \cite{CDrot}.  We  use  Newton's equation of motion, $m_i\ddot{\mathbf{r}}_i=\sum_j[\mathbf{F}_{ij}^\mathrm{dis}+\mathbf{F}_{ij}^\mathrm{el}]$, where the sum is over all particles $j$ in contact with $i$, and  the mass $m_i$ is taken  proportional to  the particle volume $v_i$.
The rotation of  particle $i$ is governed by, $\mathbf{I}_i\cdot\dot{\boldsymbol{\omega}}_i=\sum_j\mathbf{s}_{ij}\times [\mathbf{F}_{ij}^\mathrm{dis}+\mathbf{F}_{ij}^\mathrm{el}$], where $\mathbf{I}_i$ is the moment of inertia tensor of   $i$.  

We take  as unit of length $D_s\equiv\sqrt[3]{a_{1s}a_{2s}^2}=1$,  unit of energy $k_e=1$,  unit of mass $m_s=1$ and  unit of time  $t_0=D_s\sqrt{m_s/k_e}=1$.  Collision elasticity is measured by $Q=\sqrt{m_s k_e}/(k_d D_s)=2$, which would be the quality factor of a corresponding damped oscillator.  We integrate the equations of motion using a modified velocity Verlet algorithm \cite{CDrot} with  step size $\Delta t/t_0=0.05$.
We use $N=1024$ particles. 

{\bf Results:}  In this work we focus on the orientational order and tumbling of  particles, rather than  rheology.  To measure  nematic ordering we compute the  tensor \cite{Borzsonyi1},
\begin{equation}
\langle T_{\mu\nu}\rangle=\left\langle\frac{d}{(d-1)N}\sum_{i=1}^N\left[\mathbf{\hat\ell}_{i\mu}\mathbf{\hat\ell}_{i\nu}-\frac{1}{d}\delta_{\mu\nu}\right]\right\rangle,
\label{eT}
\end{equation}
where $\boldsymbol{\hat\ell}_i$ is a unit vector  along the spine of the spherocylinder or the major axis of the ellipsoid,  $\mu$ and $\nu$ denote  spatial  components,  $d=2,3$ is the spatial dimension, and $\langle\dots\rangle$ denotes an average over configurations in the sheared ensemble.
The largest eigenvalue of $\langle T_{\mu\nu}\rangle$ is the magnitude of the nematic order parameter $S_2$. The corresponding eigenvector $
\boldsymbol{\hat\ell}_2$  gives the orientation of the nematic director, which by symmetry  lies in the $xy$ plane; $\theta_2$ is the angle of $
\boldsymbol{\hat\ell}_2$  with respect to the flow direction $\mathbf{\hat x}$, and  $\mathbf{S}_2=S_2 \boldsymbol{\hat\ell}_2$.

\begin{figure}
\centering
\includegraphics[width=3.5in]{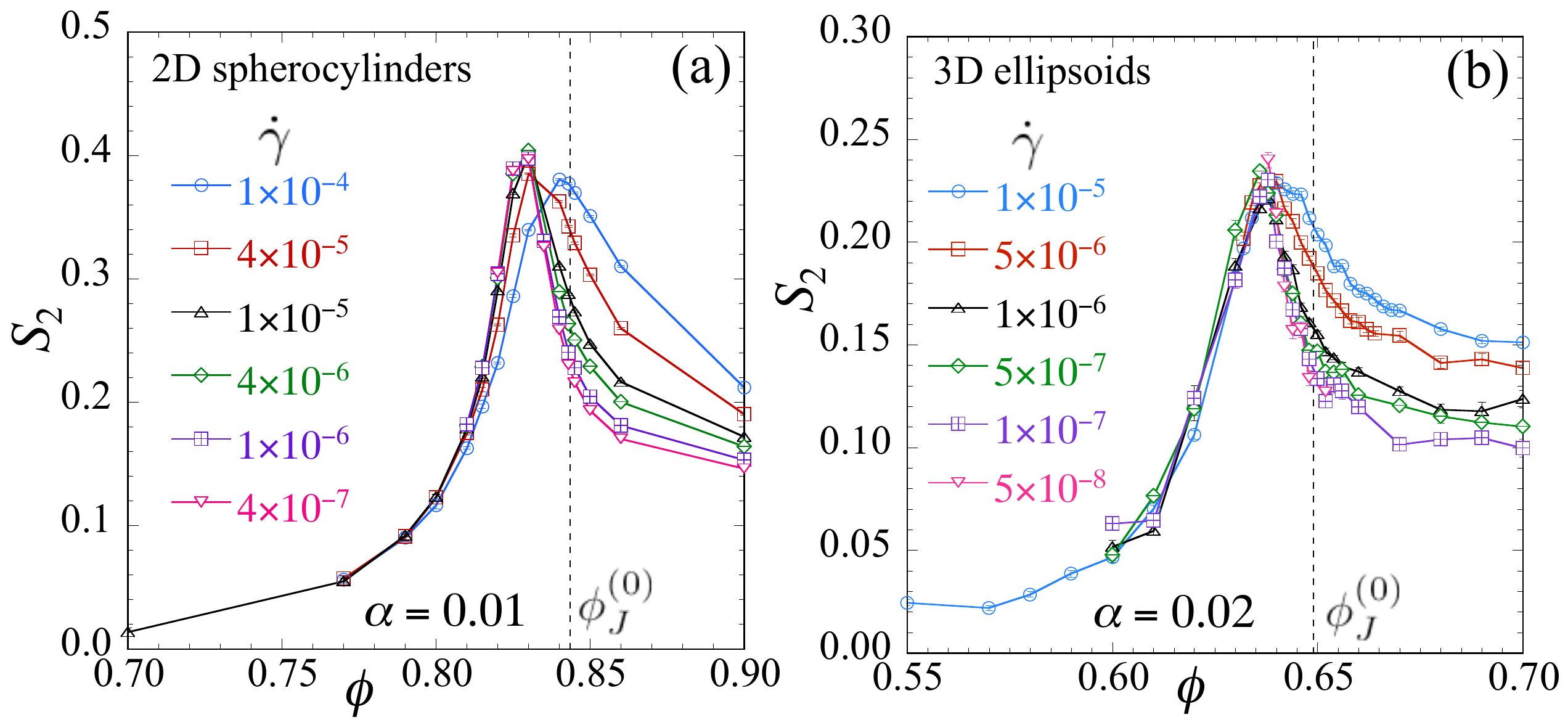}
\caption{Nematic order parameter $S_2$ vs packing  $\phi$ at different shear strain rates $\dot\gamma$.  (a) 2D spherocylinders at asphericity $\alpha=0.01$, (b) 3D ellipsoids at $\alpha=0.02$.  Vertical dashed lines locate the jamming transition of $\alpha=0$ spherical particles, $\phi_J^{(0)}=0.8433$ for 2D \cite{OT1,OT2,OT3} and 0.649 for 3D \cite{Olsson}.
}
\label{f1}
\end{figure}

In Fig.~\ref{f1} we plot $S_2$ vs  $\phi$ for  particles of fixed small asphericity $\alpha$, at different shear  rates $\dot\gamma$: (a) 
2D spherocylinders at $\alpha=0.01$, (b) 3D ellipsoids at $\alpha=0.02$. Both cases show  similar  behavior.  In contrast to previous works \cite{Campbell,Guo1,Guo2} that found increasing orientational order  with increasing $\phi$, here we find a non-monotonic $S_2$ \cite{Trulsson} with  peak value $S_{2\,\mathrm{max}}$  at a $\phi_\mathrm{max}$ slightly below the $\alpha=0$ jamming transition at $\phi_J^{(0)}$.  As $\dot\gamma$ decreases, the values of $S_2$ approach a common limiting curve \cite{Borzsonyi1,Borzsonyi2}; above $\phi_J^{(0)}$ nematic order $S_2$ stays finite, but there is a stronger $\dot\gamma$ dependence.

\begin{figure}
\centering
\includegraphics[width=3.5in]{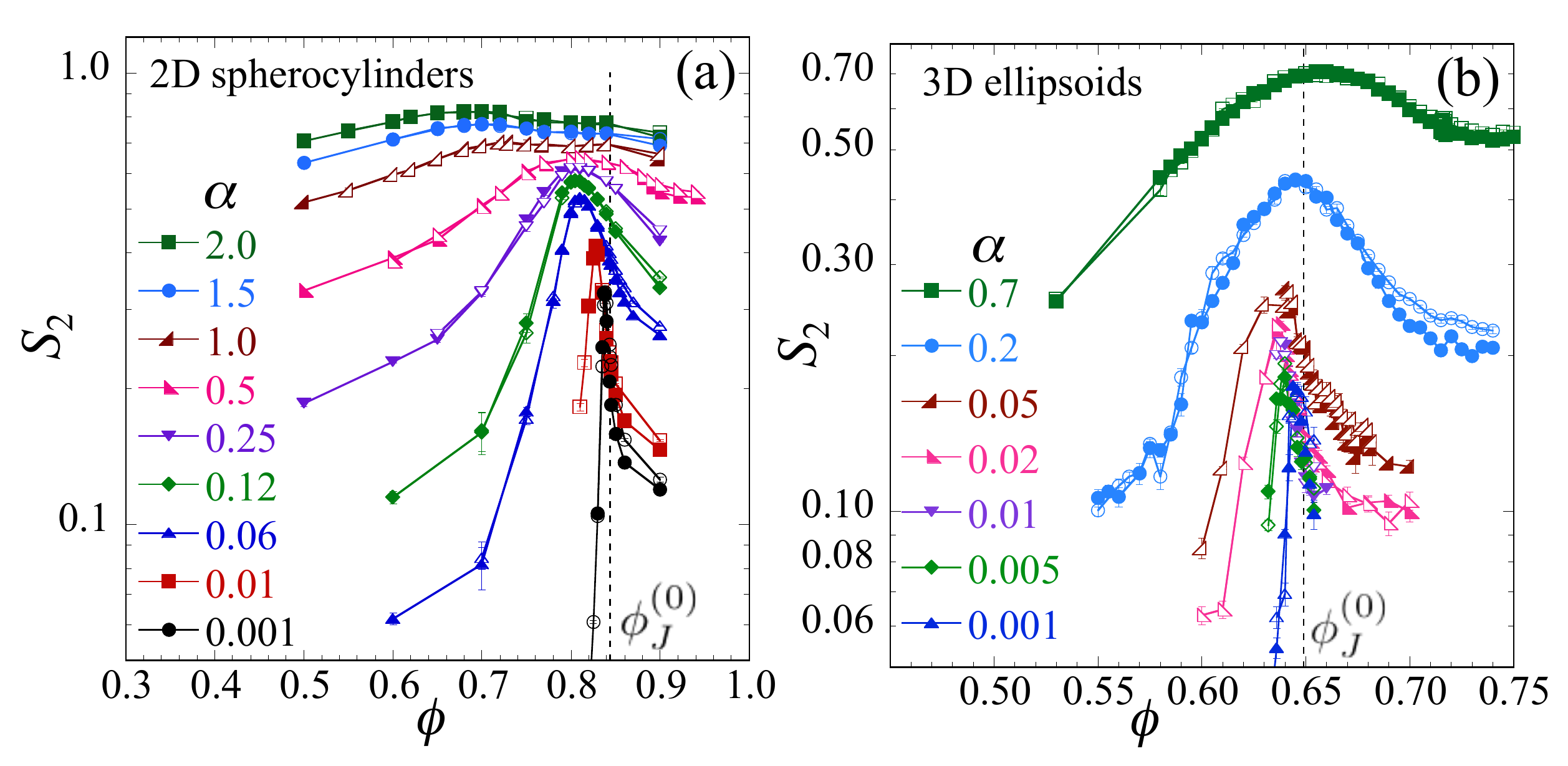}
\caption{Nematic order parameter $S_2$ for (a) 2D spherocylinders and (b) 3D ellipsoids vs packing  $\phi$ for different  asphericities $\alpha$, at  two different small strain rates $\dot\gamma_1$ (solid symbols) $< \dot\gamma_2$ (open symbols), see Table~\ref{tab1} for values.  Vertical dashed lines locate the jamming  $\phi_J^{(0)}$ of  spherical particles.
}
\label{f2}
\end{figure}

In Fig.~\ref{f2} we plot $S_2$ vs $\phi$ for a range of  $\alpha$, showing results for both a  smaller  $\dot\gamma_1$ (solid symbols) and a larger $\dot\gamma_2$ (open symbols); see  Table~\ref{tab1} for values.  In each case $\dot\gamma_1$ and $\dot\gamma_2$ are sufficiently small that $S_2$ shows no noticeable $\dot\gamma$ dependence for $\phi$ up to and slightly beyond the peak at $\phi_\mathrm{max}$, however some small $\dot\gamma$ dependence remains at the highest $\phi$. What is remarkable  is that the orientational ordering $S_{2\,\mathrm{max}}$ remains quite sizable even for particles  close to spherical with $\alpha=0.001$.

\begin{table}[h!]
\caption{Strain rate values used for data in Figs.~\ref{f2} and \ref{f3}}
\begin{center}
\begin{tabular}{|c|c|c|c|c|c|}
\hline
2D: $\alpha$ & $\dot\gamma_1$ & $\dot\gamma_2$ & 3D: $\alpha$ & $\dot\gamma_1$ & $\dot\gamma_2$ \\
\hline
0.001 & $1\times 10^{-7}$ & $4\times10^{-7}$ & $\alpha\le 0.02$ & $1\times 10^{-7}$ & $2\times 10^{-7}$\\
0.01 & $4\times 10^{-7}$ & $1\times10^{-6}$ & 0.05 & $5\times 10^{-7}$ & $1\times 10^{-6}$ \\
$\alpha\ge0.06$ & $1\times 10^{-5}$ & $4\times 10^{-5}$ & $0.2$ & $2\times 10^{-6}$ & $5\times 10^{-6}$\\
& & & 0.7 & $5\times 10^{-6}$ & $1\times 10^{-5}$\\
\hline
\end{tabular}
\end{center}
\label{tab1}
\end{table}

Fig.~\ref{f2} shows  $S_2$ averaged over the steady state  ensemble.  We have also computed the instantaneous   $S_2(\gamma)$ and $\theta_2(\gamma)$ as  functions of the system strain $\gamma=\dot\gamma t$.  We find that near and above the peak at $\phi_\mathrm{max}$, $\mathbf{S}_2(\gamma)$ shows  random fluctuations about a well defined average; there is no macroscopically coherent tumbling of particles \cite{SupMat2}.  One can still ask   if individual particles  tumble incoherently  \cite{Borzsonyi2,Wegner}, or whether they are orientationally locked into small fluctuations about the nematic director $\boldsymbol{\hat\ell}_2$.  We find the former to be the case.  

\begin{figure}
\centering
\includegraphics[width=3.5in]{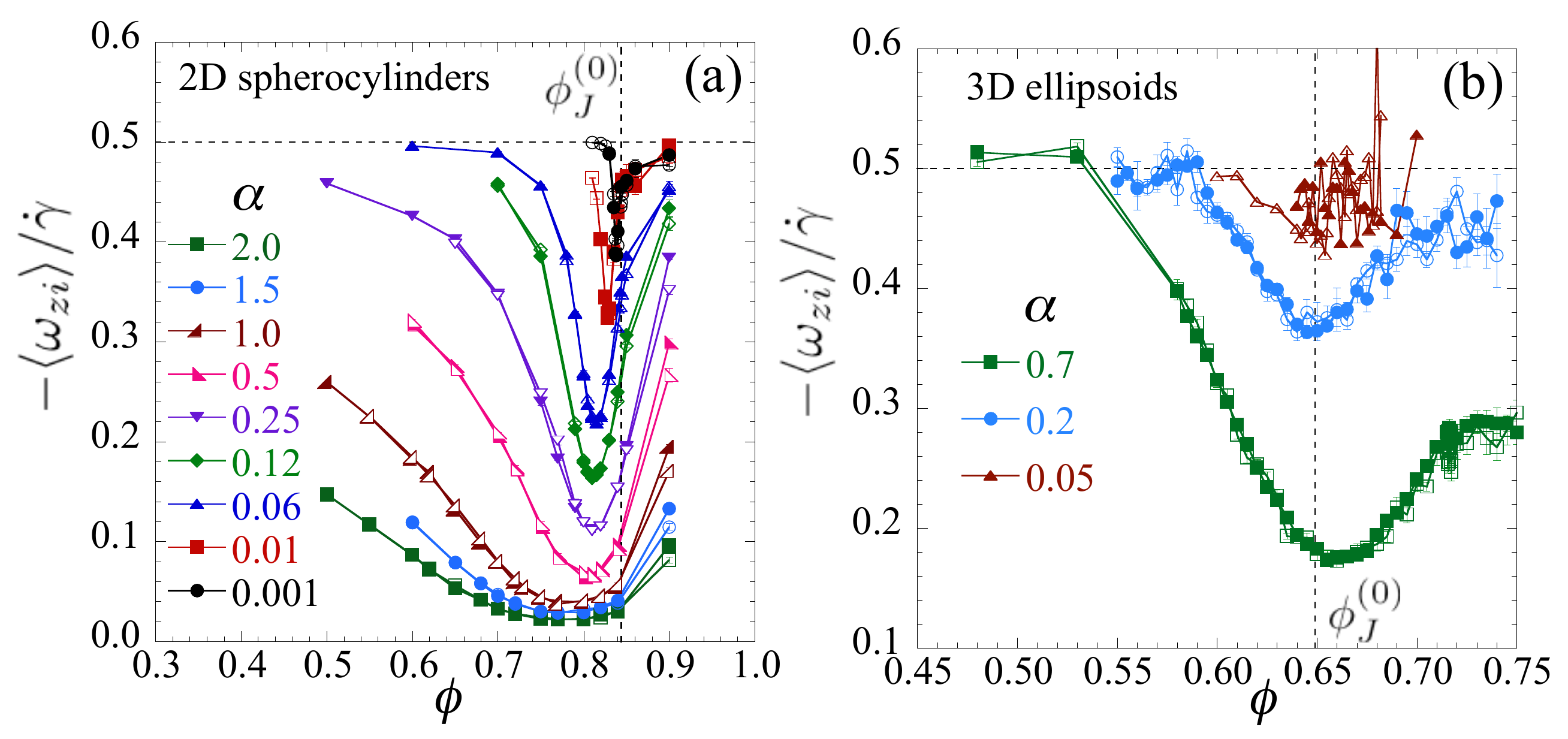}
\caption{Component of average particle angular velocity in the direction of the system vorticity, scaled by strain rate, $-\langle\omega_{zi}\rangle/\dot\gamma$ for (a) 2D spherocylinders and (b) 3D ellipsoids vs packing  $\phi$ for different    asphericities $\alpha$, at two different small strain rates $\dot\gamma_1$ (solid symbols) $< \dot\gamma_2$ (open symbols), see Table~\ref{tab1} for values.  Vertical dashed lines locate the jamming  $\phi_J^{(0)}$ of spherical particles.
}
\label{f3}
\end{figure}

In Fig.~\ref{f3} 
we plot the component of the average particle angular velocity in the direction of the system vorticity,  scaled by the strain rate, $-\langle\omega_{zi}\rangle/\dot\gamma$; note, $-\langle\omega_{zi}\rangle>0$ indicates  clockwise rotation.  For 2D spherocylinders, $\omega_{zi}=\dot\theta_i$.  In each case we show results at two different strain rates $\dot\gamma_1<\dot\gamma_2$, as in Fig.~\ref{f2} (see Table~\ref{tab1}), and find little dependence on $\dot\gamma$ except for the largest $\phi$.  Comparing Figs.~\ref{f2} and \ref{f3} we find that rotation velocity is anti-correlated with orientational order; $-\langle\omega_{zi}\rangle/\dot\gamma$ is non-monotonic in $\phi$ and is smallest when $S_2$ is largest.  
For small  but finite $\alpha$,   $-\langle\omega_{zi}\rangle/\dot\gamma$ approaches the spherical limit of 1/2 at small $\phi$, but shows a significant dip below 1/2 at $\phi_\mathrm{max}$.  For 2D spherocylinders this dip remains sizable even for  very small  $\alpha=0.001$.  For 3D ellipsoids we cannot get accurate results at similar small values of $\alpha$, but  Fig.~\ref{f3}b shows that the trends appear to be the same.  
We conclude that particles continue to rotate, with  finite $\langle\omega_{zi}\rangle/\dot\gamma$, even above jamming.

\begin{figure}
\centering
\includegraphics[width=3.5in]{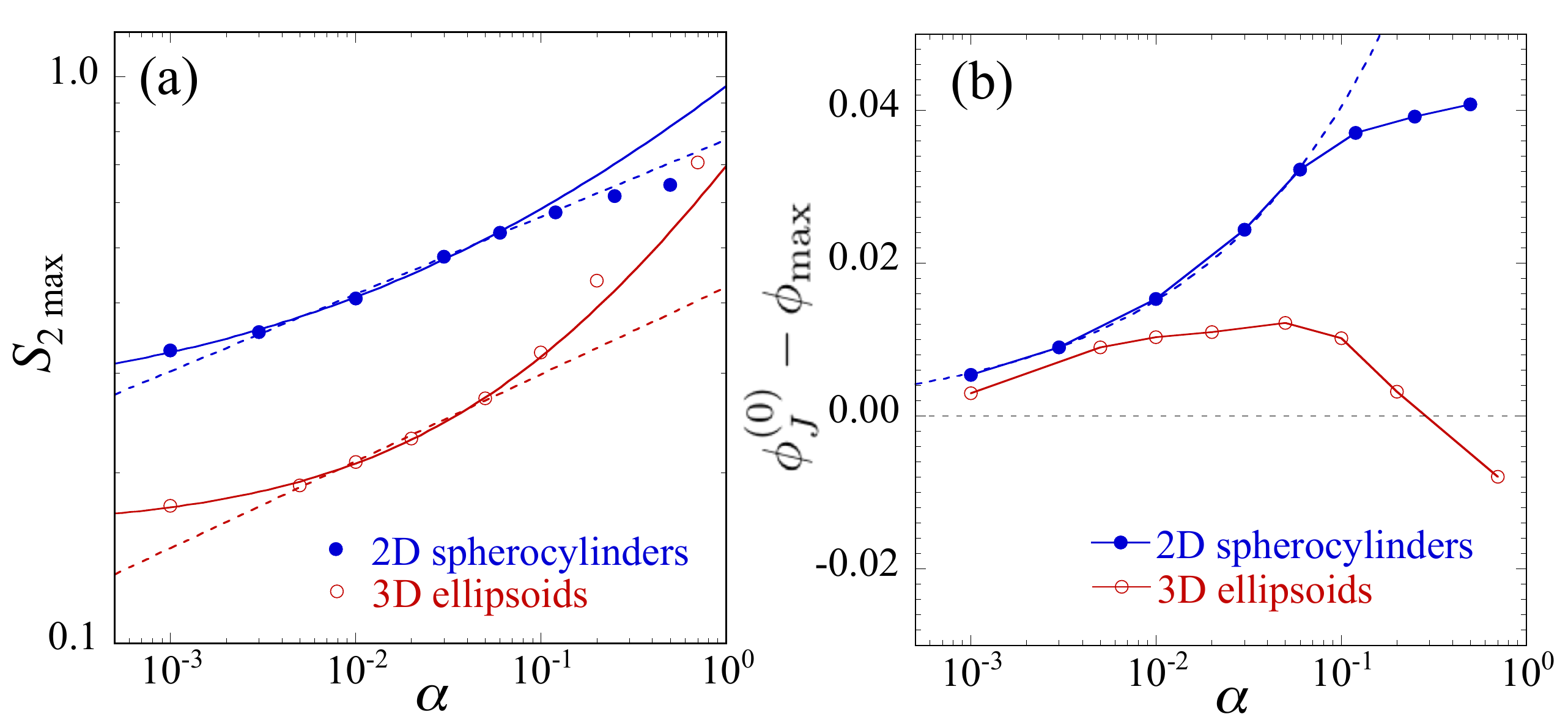}
\caption{For 2D spherocylinders and 3D ellipsoids: (a) $S_{2\,\mathrm{max}}$ vs $\alpha$.  Solid lines are fits to $S_0+c\alpha^\beta$, using the five smallest $\alpha$ points.  Dropping the  point at $\alpha=0.001$, dashed lines show power law fits.
(b) $\phi_J^{(0)}-\phi_\mathrm{max}$ vs $\alpha$, with $\phi_J^{(0)}$ the $\alpha=0$ jamming point. Solid lines connect the data points; dashed line for the 2D spherocylinders is a power law fit to the five smallest $\alpha$ points. 
}
\label{f4}
\end{figure}

\begin{figure}
\centering
\includegraphics[width=3.5in]{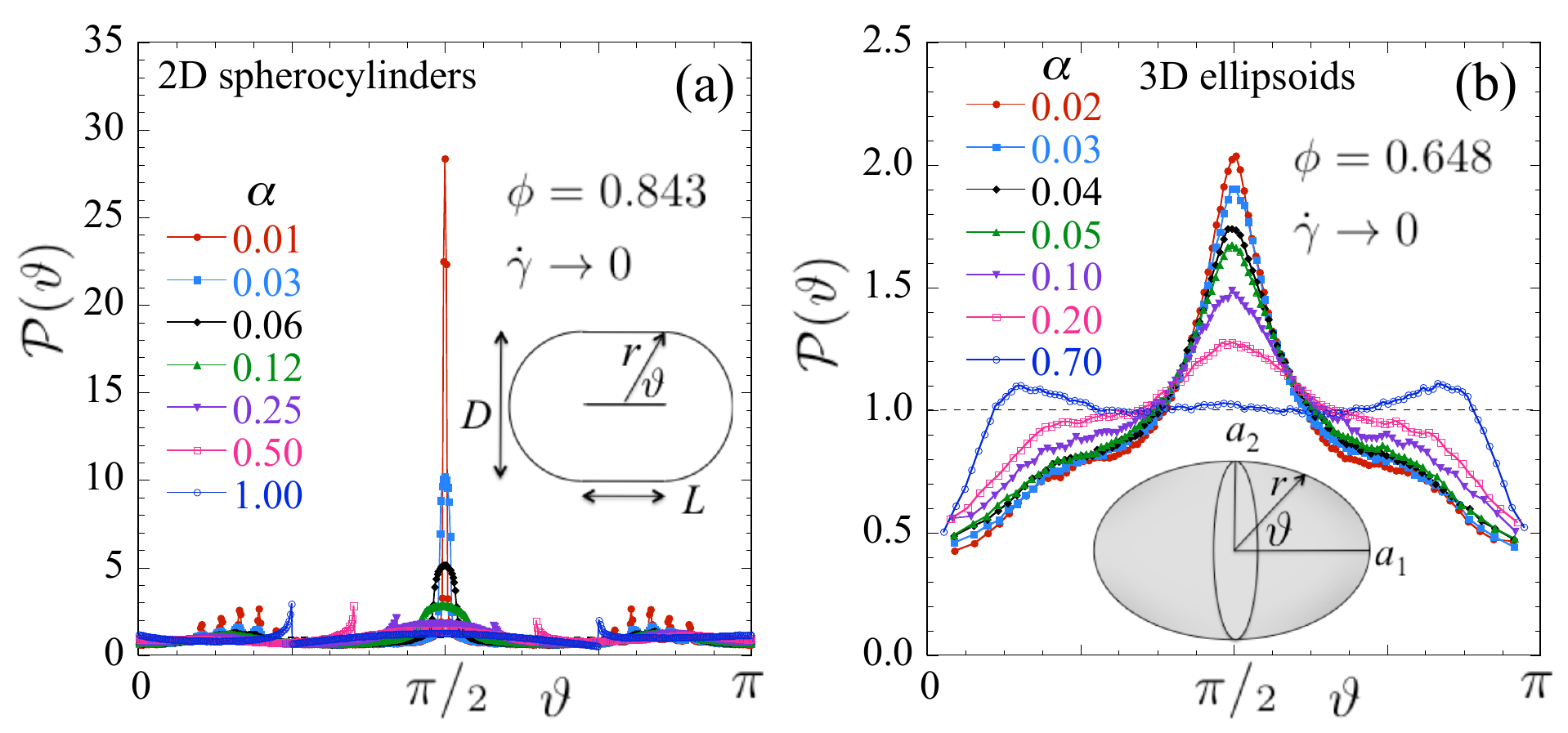}
\caption{Probability  $\mathcal{P}(\vartheta)$ for a particle to have a contact at polar angle $\vartheta$ on its surface,  for  different asphericities $\alpha$ at fixed $\phi$ near $\phi_J^{(0)}$: (a) 2D spherocylinders at $\phi=0.843$ and (b) 3D ellipsoids at $\phi=0.648$, for sufficiently small $\dot\gamma$  that $\mathcal{P}(\vartheta)$ becomes independent of $\dot\gamma$.  
In (a) the sharp peaks near $\vartheta=\pi/6$ and $5\pi/6$ are shadow effects from particles in contact  at $\vartheta=\pi/2$.
}.
\label{f5}
\end{figure}

Returning to the nematic ordering, in Fig.~\ref{f4}a we plot $S_{2\,\mathrm{max}}$ vs $\alpha$ for both 2D spherocylinders and 3D ellipsoids.  Solid lines are fits to the empirical form $S_{2\,\mathrm{max}}=S_0+c\alpha^\beta$, using the five smallest $\alpha$ points. We find $S_0=0.25$ for 2D spherocylinders and $S_0=0.16$ for 3D ellipsoids.  
If we exclude the data point at the smallest $\alpha=0.001$, then our data would be reasonably fit (dashed lines in Fig.~\ref{f4}a) by a pure power law with  exponent $\approx 0.14$.  However, in \cite{SupMat3} we give detailed tests confirming that our data point at $\alpha=0.001$ is accurate and so should not be excluded.

In Fig.~\ref{f4}b we plot $\phi_J^{(0)}-\phi_\mathrm{max}$ vs $\alpha$, where $\phi_J^{(0)}$ is the jamming transition for spherical particles.  In both 2D and 3D  we find $\phi_J^{(0)}-\phi_\mathrm{max}\to 0$ as $\alpha\to 0$,  showing that the peak in $S_2$ approaches the jamming transition as $\alpha\to 0$.  For  2D spherocylinders we find a power law dependence, $\phi_J^{(0)}-\phi_\mathrm{max}\sim \alpha^\Delta$ with $\Delta\approx 0.43$, as illustrated by the dashed line in the figure.  For 3D ellipsoids, our data do not suggest any clear form for the small $\alpha$ behavior.  
The observations of Figs.~\ref{f2} and \ref{f4} thus lead us to conclude that, even as $\alpha\to 0$ and particles are approaching the spherical limit, a finite nematic ordering $S_2$ exists at the jamming $\phi_J^{(0)}$ and above.

To look for a microscopic signature of this singular $\alpha\to 0$ limit, we  measure the location on a particle's surface of the inter-particle contacts.   
For  2D spherocylinders we define $(r,\vartheta)$ as the radial distance and polar angle of a point on the surface with respect to the center of the particle and the direction of the spine.  We  define $\mathcal{P}(\vartheta)$ as the probability density per unit surface length to have a  contact at $\vartheta$, with normalization  $1=\mathcal{A}^{-1}\int_0^{2\pi} d\vartheta\, \sqrt{r^2 + (dr/d\vartheta)^2}\,
\mathcal{P}(\vartheta)$, with $\mathcal{A}$  the perimeter length  \cite{comment}.
For  3D ellipsoids, we define $(r,\vartheta,\varphi)$ as the spherical coordinates with respect to the major axis; $\mathcal{P}(\vartheta,\varphi)$ is the probability density per unit surface area to have a contact  at $(\vartheta, \varphi)$, with normalization $1=\mathcal{A}^{-1}\int_0^{2\pi}d\varphi\int_0^\pi d\vartheta \sin\vartheta\, r\sqrt{r^2 + (dr/d\vartheta)^2}\,\mathcal{P}(\vartheta,\varphi)$; $\mathcal{A}$ is the surface area.  
For simplicity we  consider $\mathcal{P}(\vartheta)=(2\pi)^{-1}\int_0^{2\pi}d\varphi \mathcal{P}(\vartheta,\varphi)$.  
For  a uniform probability  desnity, such as would be  for spherical particles, $\mathcal{P}(\vartheta)=1$ in both 2D and 3D.

In Fig.~\ref{f5} we plot $\mathcal{P}(\vartheta)$ vs $\vartheta$ for different asphericities $\alpha$ at fixed $\phi$ near $\phi_J^{(0)}$.   
For each $\alpha$ we use a $\dot\gamma$ sufficiently small that $\mathcal{P}(\vartheta)$ has approached its $\dot\gamma\to 0$ limiting distribution.
Unlike the uniform distribution for spheres,
 we see a sharp peak at $\vartheta=\pi/2$ whose height steadily increases as $\alpha$ decreases.  
Thus as particles become increasingly spherical, particle contacts increasingly prefer to form along the narrowest width of the particle
rather than uniformly over the particle's surface \cite{contacts}.  
The onset of this effect occurs as $\phi$ increases above the jamming $\phi_J$ \cite{SupMat4}.
We note that similar results for $\mathcal{P}(\vartheta)$ have  been reported \cite{Marschall, OHern} in static, isotropically jammed packings, but in that case there is no nematic ordering and $S_2=0$.  
One possible explanation for this difference is that it is the anisotropy of the stress in a sheared system, as manifested by directed force chains, that determines a particular direction and gives rise to a non-vanishing $S_2$.
Indeed we find that for small $\alpha$ close to and above jamming (but not well below jamming and not for larger $\alpha$), the orientation $\theta_2$ of the nematic director aligns with the minimum stress axis of the stress tensor, which is at $45^\circ$ with respect to the flow direction $\mathbf{\hat x}$.

To examine the role that stress anisotropy plays, we have carried out preliminary simulations of 2D spherocylinders under a {\em pure} shear, compressing our system  in the $\mathbf{\hat y}$ direction while expanding in the $\mathbf{\hat x}$ direction, both at  constant rate $\dot\gamma/2=5\times 10^{-7}$ so as to keep   constant area.  While simple shear creates a vorticity in the affine velocity field that drives the continuous rotation of individual particles (as in our Fig.~\ref{f3}), such vorticity is absent in pure shear; we thus find $\langle\dot\theta_i\rangle=0$, the nematic director aligns with the minimal stress axis, and the magnitude $S_2$ is large at small $\phi$, monotonically {\em decreasing} as $\phi$ increases. We find $S_2$ from pure shear and simple shear qualitatively agree {\em only} when one is close to or above the jamming $\phi_J$, where  behavior is likely dominated by extended force chains that restrict particle alignment.  For small $\alpha$, pure shear and simple shear  differ most at lower $\phi$: for pure shear particles decay to a fixed orientation  giving large $S_2$ and $\theta_2=0$, while for simple shear particles continuously rotate averaging out to a small $S_2$; as $\phi$ increases, elastic collisions increase, the rotation slows and becomes more non-uniform, and $S_2$ increases.
The non-monotonic behavior of $S_2$ with a peak at $\phi_\mathrm{max}$ is thus a direct consequence of the rotational drive that is present in simple shear but absent in pure shear.  See further details in \cite{SupMat5}.

To conclude, we have considered the athermal uniform shearing of bidisperse, aspherical particles in 2D and 3D.  
A finite particle asphericity $\alpha$ breaks rotational symmetry, and  as in earlier works  \cite{Campbell,Guo1,Guo2,Borzsonyi1,Borzsonyi2,Wegner,Wegner2,Nagy,Trulsson} we find a finite nematic  ordering $S_2$.  
However one would naively expect that $S_2\to 0$ as the symmetry breaking parameter $\alpha\to 0$.  In contrast, here we show that $S_2$ remains finite at jamming and above even as $\alpha\to 0$.   This may be viewed in analogy with an Ising model, where the magnetization $m$ stays finite even as the ordering magnetic field $h\to 0$ for $T<T_c$.  However there are two significant differences: (i) In the Ising model with $h\to0$, one has $m\to 0$ as $T\to T_c$ from below, while here as $\alpha\to 0$ we find $S_2$ stays finite as $\phi\to \phi_J^{(0)}$ from above; (ii) ordering in the Ising model arises from a microscopic spin-spin interaction  that prefers  alignment even when $h=0$, while here the microscopic interaction that prefers alignment of the particle major axes would naively seem to vanish as $\alpha\to 0$ and the particles become spherical (though the behavior of $\mathcal{P}(\vartheta)$ suggests that a local ordering interaction may indeed persist even as $\alpha\to 0$).  

It would be interesting to see how robust this effect is to the introduction of additional sources of fluctuation, such as a polydispersity in $\alpha$ \cite{poly}, or the presence of thermal effects.  We  leave such questions to future research.

Simulations were  carried out on resources of the Center for Integrated Research Computing at the University of Rochester and of the Swedish National Infrastructure for Computing (SNIC) at  HPC2N.  This work was supported in part by National Science Foundation Grant No. CBET-1435861.

\bibliographystyle{apsrev4-1}




\newpage

\begin{widetext}

\centerline{\bf{\large Orientational Ordering in Athermally Sheared, Aspherical, Frictionless Particles}}
\vskip 8pt
\centerline{\bf{\large Supplemental Material}}
\vskip 8pt

\end{widetext}

%
\setcounter{figure}{0}
\setcounter{equation}{0}
\renewcommand{\theequation}{SM-\arabic{equation}}
\renewcommand{\thefigure}{SM-\arabic{figure}}

In this Supplemental Material we provide further details and tests to demonstrate the correctness of our simulations.  
In Sec.~\ref{valid} we demonstrate the validity of our results for the nematic order parameter $S_2$ at our smallest asphericity, $\alpha=0.001$, which is key to our conclusion that $S_{2\,\mathrm{max}}$ stays finite as $\alpha\to 0$.
In Sec.~\ref{flow} we discuss translational correlations in our system and demonstrate that there is no smectic ordering into well defined flowing layers.
In Sec.~\ref{contact} we show that the onset for the effect that particle contacts to prefer to lie on the narrowest width of the particles, takes place as the packing $\phi$ increases through the jamming transition.
In Sec.~\ref{pure} we consider the effect of a pure shear deformation on 2D spherocylinders, and contrast with our main results for simple shear.
In Sec.~\ref{deltaij} we provide details on how we determine when two particles are in contact, and compute the corresponding overlap parameters.

\section{Validity of results at $\alpha=0.001$}
\label{valid}

Our argument in the main text, that $\lim_{\alpha\to 0}[S_{2\,\mathrm{max}}]=S_0$ is finite, relied on the assertion that our data at the smallest $\alpha=0.001$ are reliable.  In order to argue conversely, i.e., that $S_{2\,\mathrm{max}}$  vanishes as a power law as $\alpha\to 0$, we would have to believe that the value of $S_{2\,\mathrm{max}}$ at $\alpha=0.001$ that is  reported in Fig.~4a of the main text is, by some artifact of our simulations, larger than it should be.

Here we provide several tests to support our claim that our data point at $\alpha=0.001$ is indeed correct.  Since our simulations for 2D spherocylinders are considerably less time consuming than for 3D ellipsoids, we can make more exacting tests for that case.  Hence, here we restrict ourselves to 2D spherocylinders.

\subsection{Dependence on Shear Strain Rate}

\begin{figure}[h!]
\centering
\includegraphics[width=3.5in]{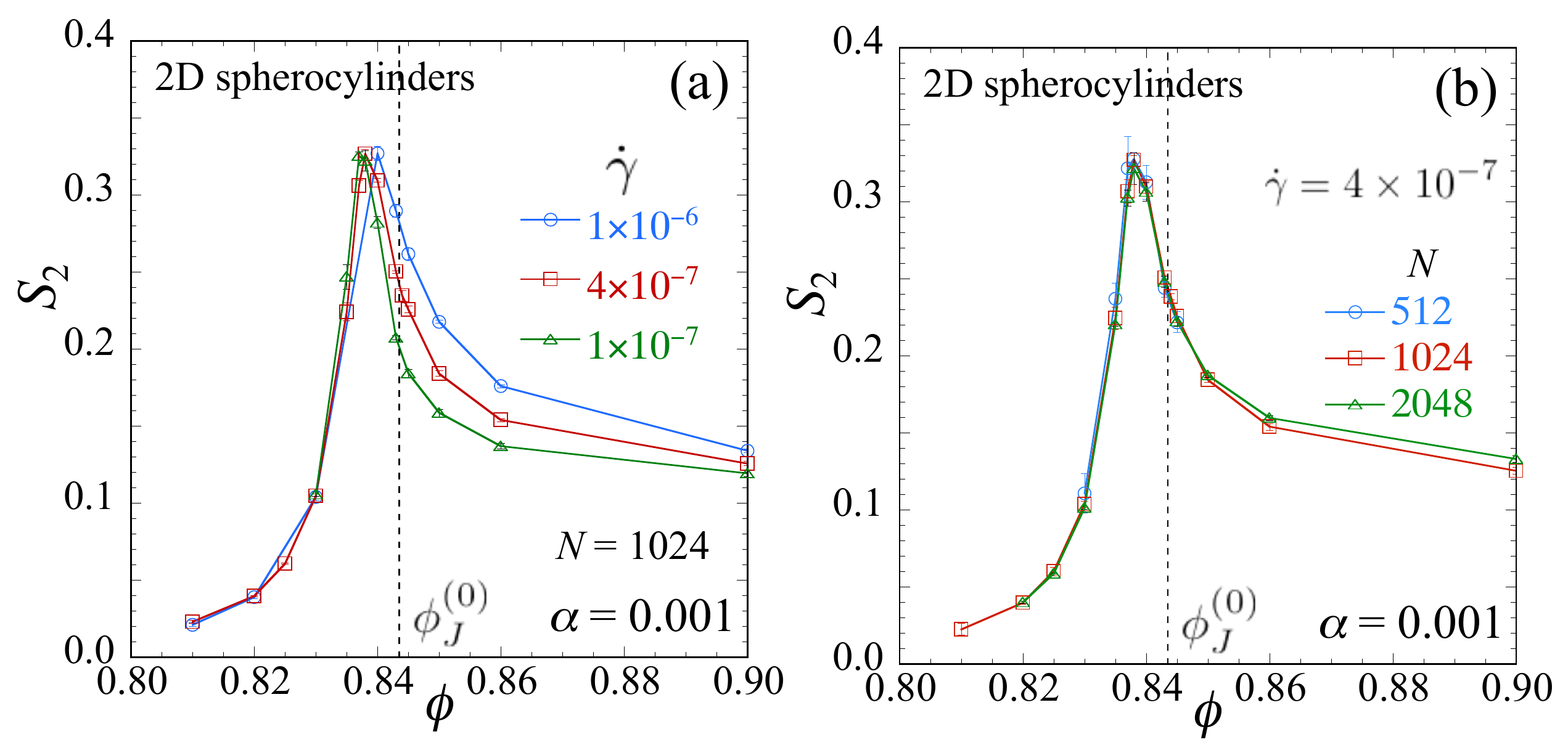}
\caption{Nematic order parameter $S_2$ for $\alpha=0.001$ vs packing fraction $\phi$ for (a) three different shear strain rates $\dot\gamma$ with $N=1024$ particles, and  (b) systems with different numbers of particles $N$ for $\dot\gamma=4\times 10^{-7}$.  Vertical dashed lines locate the jamming transition of $\alpha=0$ spherical particles, $\phi_J^{(0)}=0.8433$.
}
\label{fSM1}
\end{figure}

As shown in Fig.~1 of the main text, the nematic order parameter $S_2$ depends on both packing fraction $\phi$ and shear strain rate $\dot\gamma$.  However at each $\phi$, $S_2$ approaches a limiting value as $\dot\gamma$ decreases.  Here we wish to confirm that we have simulated at small enough $\dot\gamma$ so that the peak value $S_{2\,\mathrm{max}}$ which we find for $\alpha=0.001$ has reached this $\dot\gamma\to 0$ limit.  In Fig.~\ref{fSM1}a we plot $S_2$ vs $\phi$ for our three smallest strain rates $\dot\gamma$, using a system with $N=1024$ particles.  Just as was found in Fig.~1 of the main text for a larger $\alpha$, here we see $\dot\gamma$ dependence remaining on the large $\phi$ side of the peak in $S_2$, however there is no $\dot\gamma$ dependence on the low $\phi$ side up to, and including, the peak value.  
Thus our results of Fig.~\ref{fSM1}a clearly argue that the value of $S_{2\,\mathrm{max}}$ will not decrease if $\dot\gamma$ were made even smaller.

\subsection{Dependence on System Size}

As one approaches the jamming transition, a correlation length diverges.  If one is too close to the jamming transition, finite size effects may become important when the correlation length becomes larger than the length of the system.  We thus wish to check that our value of $S_{2\,\mathrm{max}}$ for $\alpha=0.001$ is not affected by such possible finite system size effects.  Our results in the main text are for systems with $N=1024$ particles.   In Fig.~\ref{fSM1}b we plot $S_2$ vs $\phi$ at the small strain rate $\dot\gamma=4\times 10^{-7}$, using three different systems sizes with numbers of particles $N=512, 1024$ and 2048.  While there is a small dependence on $N$ seen at the largest $\phi$, there is no dependence on $N$ at lower $\phi$ up to and including the peak value.  Our value of $S_{2\,\mathrm{max}}$ for $\alpha=0.001$  thus does not suffer from finite size effects. 

\subsection{Dependence on Integration Time Step}

\begin{figure}[h!]
\centering
\includegraphics[width=2.0in]{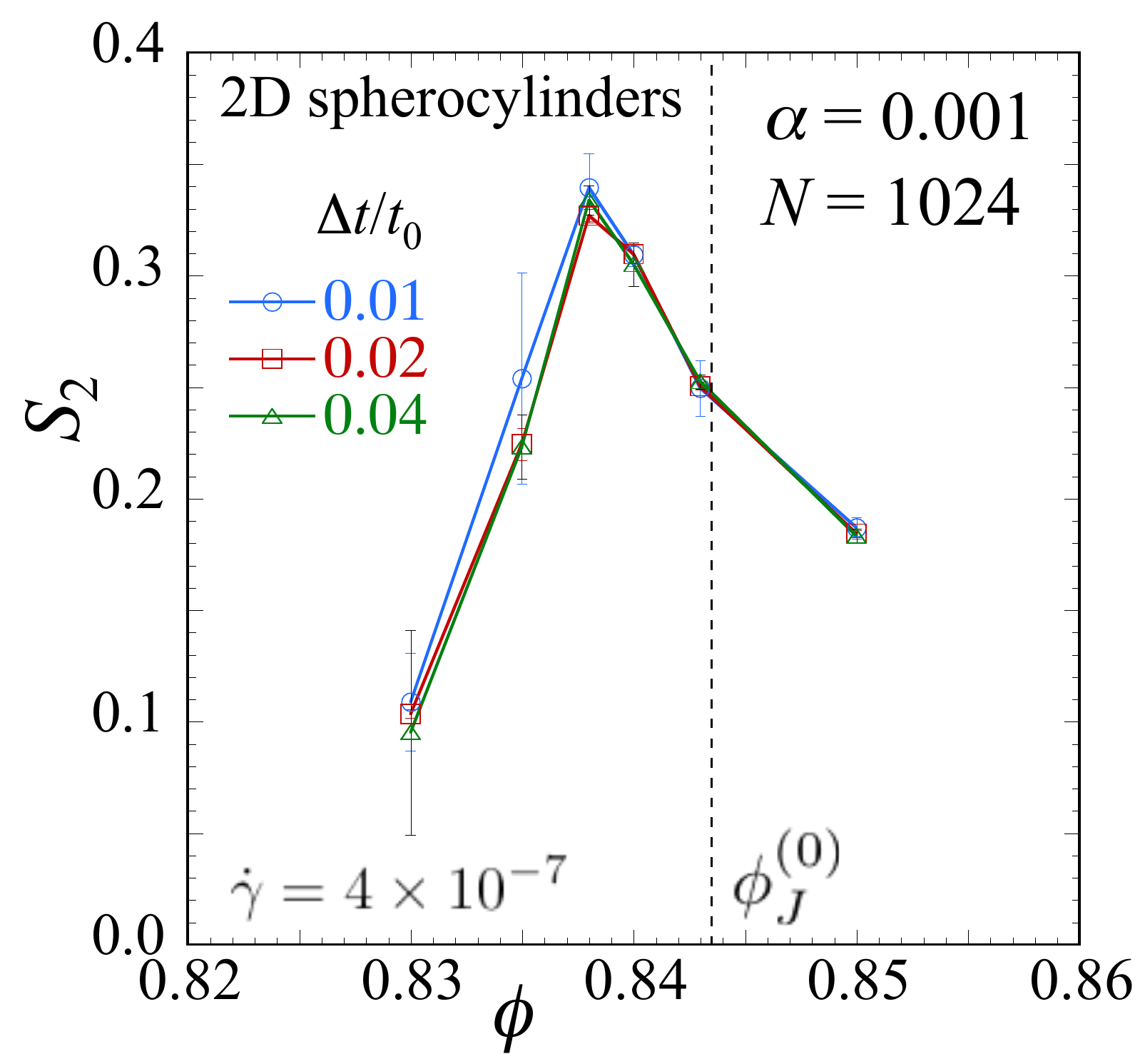}
\caption{Nematic order parameter $S_2$ for $\alpha=0.001$ vs packing fraction $\phi$ using  different integration time steps  $\Delta t/t_0=0.01, 0.02$ and 0.04.  The system is sheared at a strain rate $\dot\gamma=4\times 10^{-7}$ and has $N=1024$ particles. Vertical dashed line locates the jamming transition of $\alpha=0$ spherical particles, $\phi_J^{(0)}=0.8433$.
}
\label{fSM2}
\end{figure}

We should also check if there is any dependence of our results on the size of the finite numerical integration step $\Delta t$.  Our results in the main text used a value $\Delta t=0.02 t_0$ with $t_0=D_s^2k_d/k_e$ the unit of time.  In Fig.~\ref{fSM2} we plot $S_2$ vs $\phi$ at the small strain rate $\dot\gamma=4 \times 10^{-7}$, for $\alpha=0.001$, using three different values of the time step $\Delta t/t_0=0.01, 0.02$ and $0.04$.  We see that any differences between the data from the three different $\Delta t$ are within the estimated statistical error.  We conclude that our integration step of $\Delta t/t_0=0.02$ is small enough to accurately determine $S_{2\,\mathrm{max}}$ for $\alpha=0.001$.  

\subsection{Equilibration}

Finally we demonstrate that  the rotational degrees of freedom in our system are well equilibrated for our simulations at $\alpha=0.001$.  When $\alpha$ is small, the small moment arms of the collisional  forces result in small torques, and, depending on the particle density, it can require long shear strains for the rotational degrees of freedom of the system to equilibrate to the proper steady state.  

Let us define $S_2(\gamma)$ as the  magnitude of the instantaneous nematic order parameter of the individual configuration of the system after it has sheared a strain $\gamma=\dot\gamma t$.  
For an initial configuration of  randomly oriented particles, a system with a finite number of particles $N$ will have some initial value $S_2(0)$.
At low densities, where torque inducing collisions are rare, particles will rotate primarily under the influence of the dissipative torque.  In this case, since particles with finite $\alpha$ have a non uniform angular velocity that depends on their orientation $\theta_i$, the relative orientations of the particle spines $\boldsymbol{\hat\ell}_i$ will vary with $\gamma$ and hence so will $S_2$.  But once  the system has strained so that $\gamma=\dot\gamma T$, where $T$ is the period of rotation of an isolated particle,  the particles will  have returned to their initial orientations and $S_2(\gamma)$ will have returned to its initial value $S_2(0)$.  We thus expect to see an oscillating $S_2(\gamma)$ with period $\dot\gamma T$.  We have observed such behavior for small $\alpha$ at low densities.  
However, as the density increases the rate of collisions increases.   These collisions will perturb this oscillatory behavior until, after a sufficiently large  strain has been applied, the particle orientations have lost memory of their initial configuration.   The particle orientations  will then sample a stationary steady state distribution.   $S_2(\gamma)$ will then be constant, aside from random fluctuations that we might expect should decrease as $1/\sqrt{N}$ as the number of particles $N$  increases. 

\begin{figure}[h!]
\centering
\includegraphics[width=3.5in]{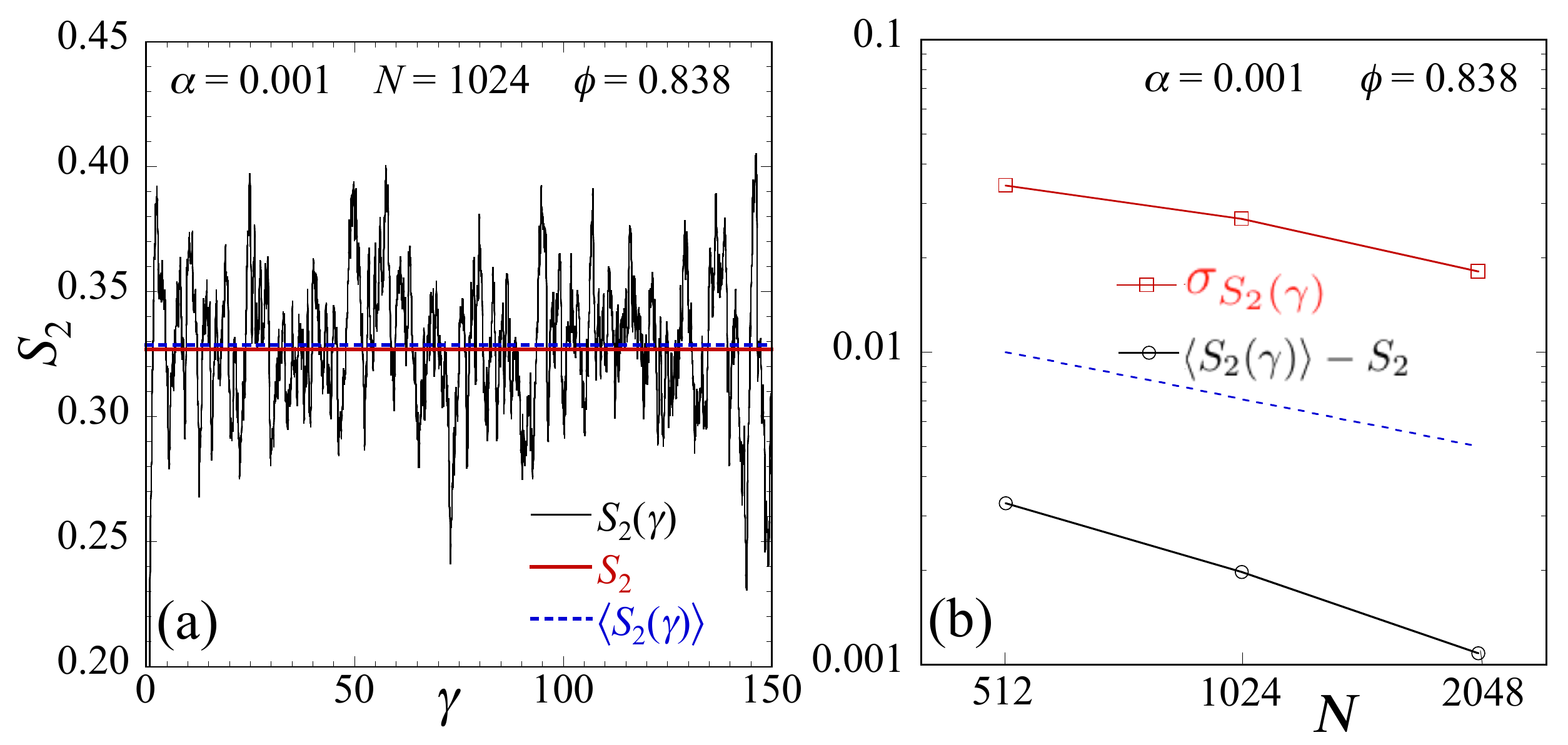}
\caption{(a) Instantaneous nematic order parameter $S_2(\gamma)$  vs shear strain  $\gamma$, for  $\alpha=0.001$ and shear strain rate $\dot\gamma=4\times 10^{-7}$ at packing fraction $\phi=0.838$    near the peak value $S_{2\,\mathrm{max}}$.  The horizontal dashed line is the average over these instantaneous values $\langle S_2(\gamma)\rangle$, while the horizontal solid line is $S_2$ as obtained from averaging the orientational ordering tensor  over the entire shearing run.
The system has $N=1024$ particles. (b) Difference $\langle S_2(\gamma)\rangle - S_2$ vs number of particles $N$, and standard deviation $\sigma_{S_2(\gamma)}$ vs $N$; the dashed line is $\sim 1/\sqrt{N}$ for comparison.
}
\label{fSM3}
\end{figure}

In Fig.~\ref{fSM3}a we plot $S_2(\gamma)$ vs $\gamma$  for a shear strain rate $\dot\gamma=4\times 10^{-7}$ at a packing $\phi=0.838$ near the peak in $S_2$, for a system with $N=1024$ particles with $\alpha=0.001$.  We see that $S_2(\gamma)$ appears, as desired, to consist of random fluctuations about a constant average.  The dashed horizontal line in Fig.~\ref{fSM3}a is the average $\langle S_2(\gamma)\rangle=(1/\Delta\gamma)\int_{\gamma_i}^{\gamma_f}d\gamma\,S_2(\gamma)$, where $\Delta\gamma=\gamma_f-\gamma_i$; we start averaging only after an initial shear strain of $\gamma_i=10$ so as to avoid any initial transients, and average up to a final $\gamma_f=150$.
The solid horizontal line represents the ensemble average $S_2$, as considered elsewhere in this work.  To be clear, $S_2(\gamma)$ is the largest eigenvalue of the orientational ordering tensor $T_{\mu\nu}(\gamma)$ as computed for the individual configuration at strain $\gamma$, while $S_2$ is the largest eigenvalue of the orientational ordering tensor $\langle T_{\mu\nu}\rangle$ averaged over the length of the shearing run from $\gamma_1$ to $\gamma_2$.  Since the relation between eigenvalue and tensor is not linear, these two averages of $S_2$ need not be equal, and in Fig.~\ref{fSM3}a we see that there is indeed a small difference.  Since the direction of the nematic director is optimized to give the largest possible $S_2$, and since the direction of the nematic director  obtained from $T_{\mu\nu}(\gamma)$ fluctuates as $\gamma$ varies from configuration to configuration (as opposed to the director obtained from $\langle T_{\mu\nu}\rangle$ which is fixed), we expect that $\langle S_2(\gamma)\rangle$ will be somewhat larger than $S_2$, and this is indeed what is observed in Fig.~\ref{fSM3}a.  In Fig.~\ref{fSM3}b we plot $\langle S_2(\gamma)\rangle - S_2$ vs $N$ and see that this difference is going to zero as $N$ increases.  In the same figure we also plot the standard deviation $\sigma_{S_2(\gamma)}= \sqrt{\langle S_2^2(\gamma)\rangle -\langle S_2(\gamma)\rangle^2}$ vs $N$ and see that it also vanishes  as $N$ increases.

\begin{figure}[h!]
\centering
\includegraphics[height=1.75in]{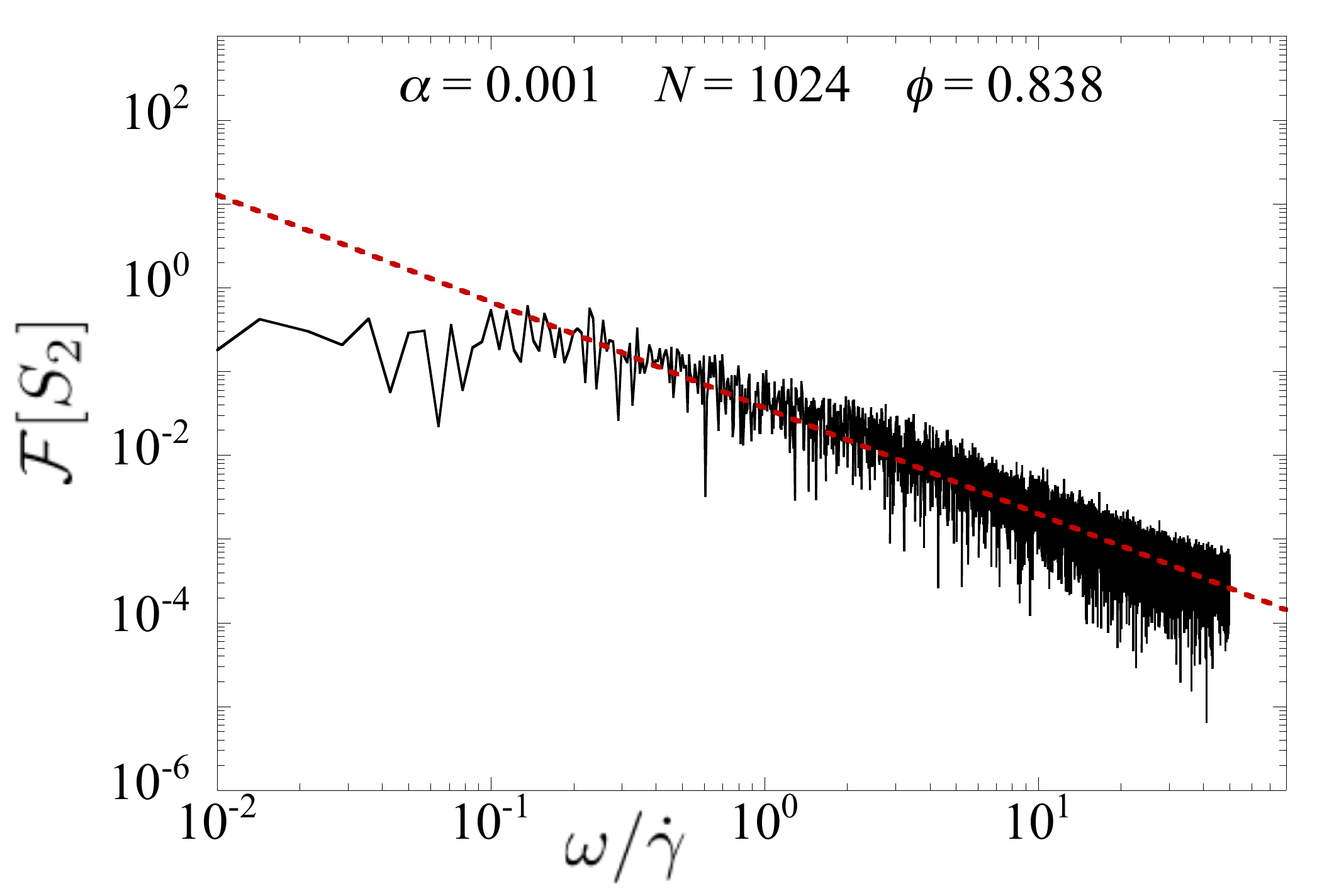}
\caption{Fourier transform of $S_2(\gamma)$, $\mathcal{F}[S_2]$,  vs dimensionless frequency $\omega/\dot\gamma$.  The high frequency tail is fit to an inverse power law (dashed line) and gives an exponent $\sim 1.3$.
}
\label{fSM4}
\end{figure}

Next, we consider the Fourier transform of $S_2(\gamma)$ in order to check that the frequency spectrum of the fluctuating noise seen in Fig.~\ref{fSM3}a is broad without any peaks that could indicate vestigial oscillations due to poor equilibration.  Since $S_2(\gamma)$ is plotted in terms of the dimensionless time $\gamma=\dot\gamma t$, in Fig.~\ref{fSM4} we plot the Fourier transform $\mathcal{F}[S_2]$ as a function of the dimensionless frequency $\omega/\dot\gamma$.  We see that the spectrum is indeed broad with no peaks.  The high frequency tail is roughly power law with an exponent $~1.3$, however that exponent changes a bit depending on the range of $\gamma$ that is used in the fit.  

\begin{figure}
\centering
\includegraphics[width=3.5in]{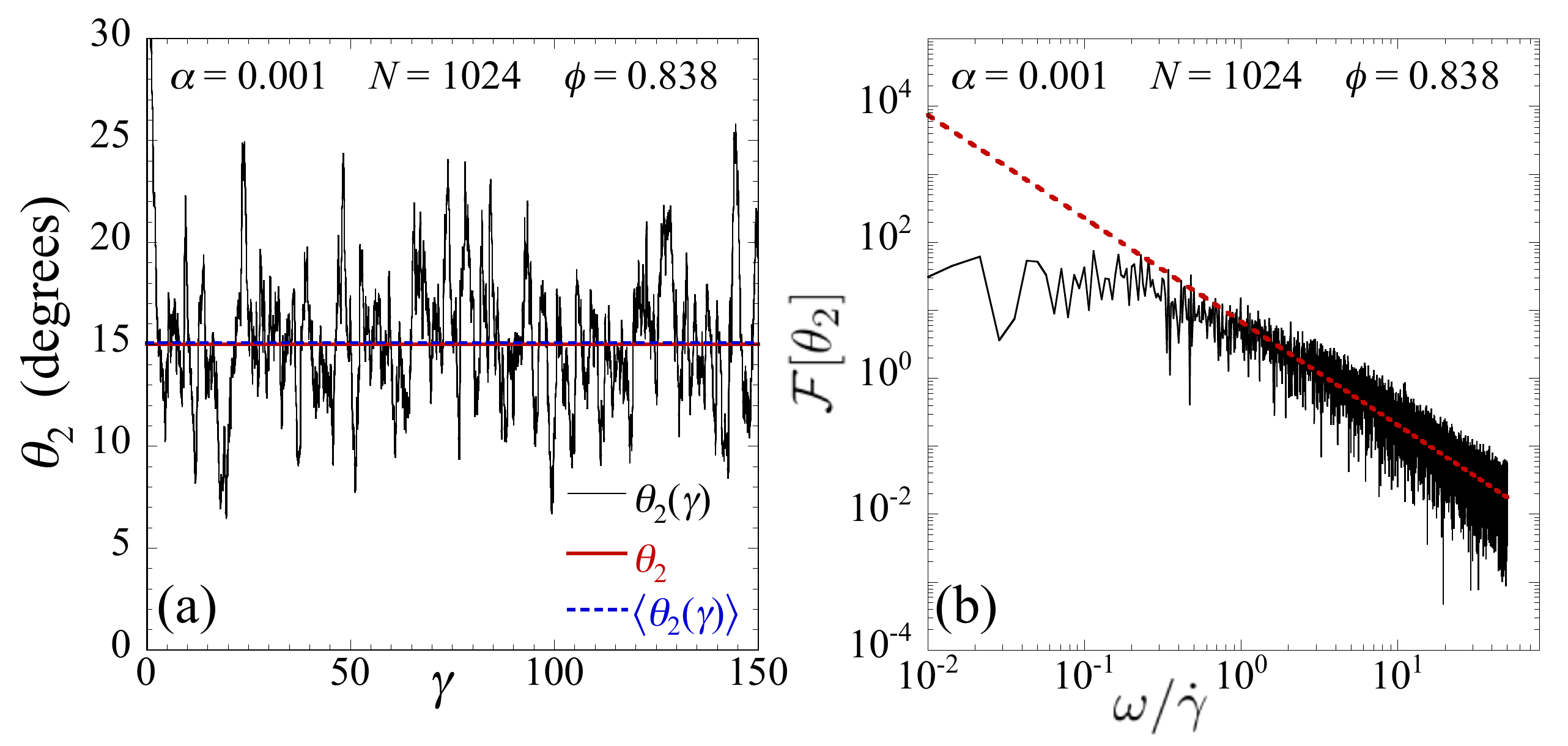}
\caption{(a) Instantaneous nematic director angle $\theta_2(\gamma)$  vs shear strain  $\gamma$, for  $\alpha=0.001$ and shear strain rate $\dot\gamma=4\times 10^{-7}$ at packing fraction $\phi=0.838$  near the peak value $S_{2\,\mathrm{max}}$.  The horizontal dashed line is the average over these instantaneous values $\langle \theta_2(\gamma)\rangle$, while the horizontal solid line is $\theta_2$ as obtained from averaging the orientational ordering tensor  over the entire shearing run.
The system has $N=1024$ particles. (b) Fourier transform of $\theta_2(\gamma)$, $\mathcal{F}[\theta_2]$, vs dimensionless frequency $\omega/\dot\gamma$.  The high frequency tail is fit to an inverse power law (dashed line) and gives an exponent $\sim 1.5$.
}
\label{fSM5}
\end{figure}

Lastly we consider a similar analysis of the orientation angle $\theta_2$ of the nematic director.  In Fig.~\ref{fSM5}a we plot the  instantaneous $\theta_2(\gamma)$ vs $\gamma$ for  the same parameters as in Fig.~\ref{fSM3}a, $\dot\gamma=4\times 10^{-7}$, $\phi=0.838$, $\alpha=0.001$.  We see what appear to be random fluctuations about a constant average value.  The dashed horizontal line is the average $\langle \theta_2(\gamma)\rangle = (1/\Delta\gamma)\int_{\gamma_i}^{\gamma_f}d\gamma\,\theta_2(\gamma)$, while the solid horizontal line is $\theta_2$ obtained from the ensemble averaged orientation tensor $\langle T_{\mu\nu}\rangle$.  In Fig.~\ref{fSM5}b we plot the Fourier transform $\mathcal{F}[\theta_2]$ vs the dimensionless frequency $\omega/\dot\gamma$.  We see a broad spectrum with a power law tail decreasing with an exponent $\sim 1.5$ (the exact value of this exponent is sensitive to the range of data used in the fit).  There are no peaks in $\mathcal{F}[\theta_2]$ to indicate any oscillatory motion, thus giving support to the assertion in the main text that, while individual particles tumble with an average angular velocity $\langle \omega_i\rangle$, there is no coherent tumbling of the nematic order parameter $\mathbf{S}_2$.  Our results in this section thus confirm that our spherocylinder simulations at $\alpha=0.001$ are indeed well equilibrated.

\section{Spatial Correlations}
\label{flow}

Since our system has finite nematic orientational order, we wish to check whether there might also be smectic translational order, with particles flowing in well defined layers oriented in the direction of the flow.
To test for this, we measure the following transverse correlation function of the particle center of mass density $n(\mathbf{r})$.
We first define,
\begin{equation}
n(y)=\dfrac{1}{\Delta y \mathcal{L}_\perp^{d-1}}\int_{y-\Delta y/2}^{y+\Delta y/2}\!\!\! dy^\prime\int_0^{\mathcal{L}_\perp}\!\!\!d\mathbf{r}_\perp \, n(y^\prime, \mathbf{r}_\perp).
\end{equation}
$n(y)$  is just he number of particles per unit volume whose center of mass lies in a layer of small width $\Delta y$ that spans the system in the orthogonal directions.
For $d=2$ dimensions, $\mathbf{r}_\perp = x\mathbf{\hat x}$ and $\mathcal{L}_\perp=\mathcal{L}_x$, the length of the system in the $\mathbf{\hat x}$ direction; for $d=3$, $\mathbf{r}_\perp=x\mathbf{\hat x}+z\mathbf{\hat z}$ and $\mathcal{L}_\perp=\mathcal{L}_x=\mathcal{L}_z$. 
We then define the correlation
\begin{equation}
C(y) = \dfrac{\mathcal{L}_\perp^{d-1}}{n\mathcal{L}_y}\int_0^{\mathcal{L}_y}\! \!dy^\prime \left[\langle n(y+y^\prime)n(y^\prime)\rangle - \langle n\rangle^2\right],
\end{equation}
where the prefactor is chosen so that $C(y)$ is independent of  the system size.  

\begin{figure}[h!]
\centering
\includegraphics[width=3.5in]{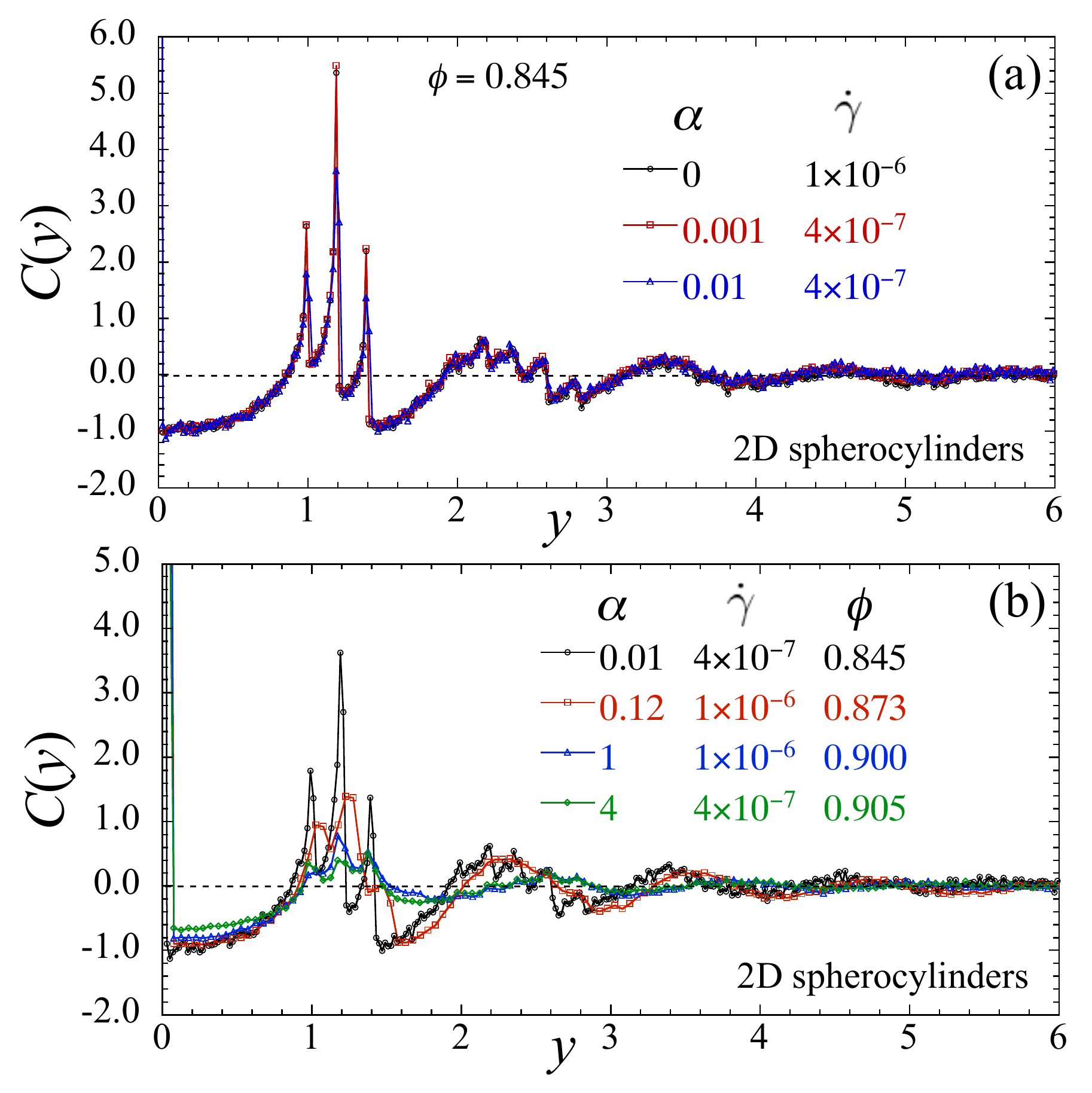}
\caption{Spatial correlations in the direction of the flow velocity gradient, $C(y)$ vs $y$, for 2D spherocylinders.  (a) Results for small $\alpha$, including $\alpha=0$, at the common value of $\phi=0.845$, just above the jamming fraction for circular disks, $\phi_J^{(0)}=0.8433$; strain rates $\dot\gamma$ are as indicated in the figure. (b) Results for larger values of $\alpha$, at low strain rates $\dot\gamma$, just above their respective jamming fractions $\phi_J(\alpha)$; values of $\dot\gamma$ and $\phi$ are indicated in the figure.
}
\label{C-vs-y-2D}
\end{figure}

We consider first the case of spherocylinders in 2D, where we  average over  large total strains $\gamma\approx 130$, thus allowing for accurate measurements of $C(y)$.  For our calculations we use a layer width $\Delta y = 0.01$ for $\alpha\le 0.01$, and $\Delta y=0.025$ for larger $\alpha$.
In Fig.~\ref{C-vs-y-2D}a we plot $C(y)$ vs $y$ for $\alpha=0$, $0.001$, and $0.01$ at $\phi=0.845$, which is just slightly above the jamming $\phi_J^{(0)}=0.8433$ for circular disks.  In each case we use the smallest $\dot\gamma$ we have simulated at each $\alpha$, i.e. $10^{-6}$, $4\times10^{-7}$ and $4\times 10^{-7}$ respectively.
We see that the $C(y)$ for these three cases are almost indistinguishable; there is nothing that signals a singular behavior as $\alpha\to 0$.  
We see sharp peaks at $y=1$, 1.2, and 1.4, which are the nearest neighbor separations for just contacting small-small, small-big, and big-big pairs.
At larger $y$ we see oscillations with a period of 1.2, the average spacing between contacting particles.  However these oscillations clearly decay to zero as $y$ increases, thus demonstrating that there is only short ranged order in the direction of the flow velocity gradient.  Fitting the heights of the larger $y$ peaks to an exponential, we find a decay length between $1$ and $2$.

In Fig.~\ref{C-vs-y-2D}b we plot $C(y)$ vs $y$ for larger values of $\alpha$, at our lowest strain rate for each case, and at a packing fraction $\phi$ that is slightly above the respective jamming fraction $\phi_J$ for each $\alpha$.  We again see similar behavior: oscillations  that decay to zero as $y$ increases.  As $\alpha$ increases, and the particles become increasingly non-spherical, the sharp peaks near $y=1$, 1.2 and 1.4 broaden and the peaks at $y>2$ shift to slightly larger values of $y$; results for $\alpha=1$ and $\alpha=4$ are nearly indistinguishable for $y>2$.
However the average spacing between peaks remains $\sim 1.2$ and the decay length remains in the range 1 to 2.  We have verified that similar behavior occurs as either $\phi$ or $\dot\gamma$ is varied.
We thus conclude that  particles do not flow in  well defined, spatially ordered, layers and so there is no smectic ordering.   

\begin{figure}[h!]
\centering
\includegraphics[width=3.5in]{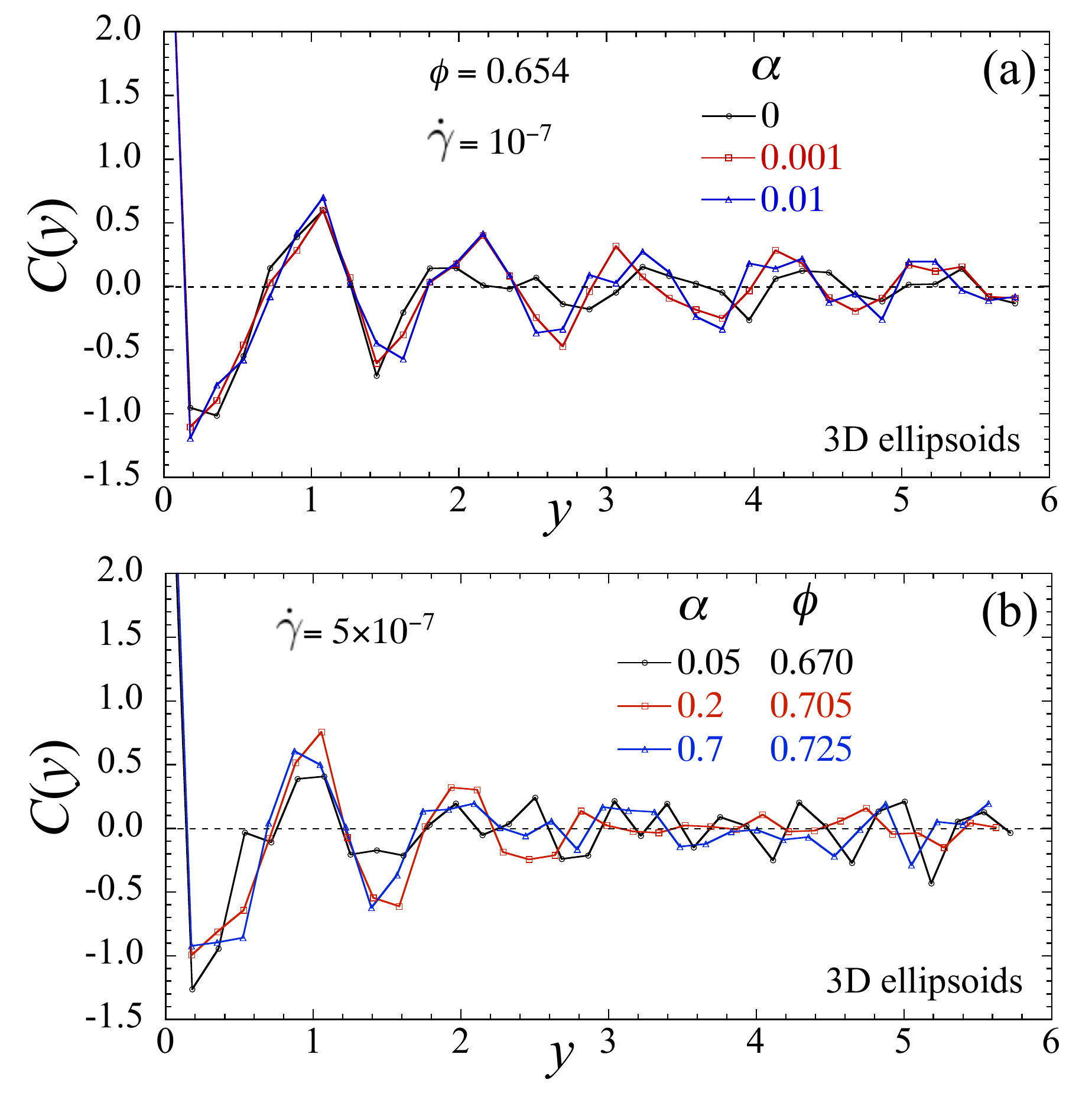}
\caption{Spatial correlations in the direction of the flow velocity gradient, $C(y)$ vs $y$, for 3D ellipsoids.  (a) Results for small $\alpha$, including $\alpha=0$, at strain rate $\dot\gamma=10^{-7}$ and the common value of $\phi=0.654$, just above the jamming fraction for spheres, $\phi_J^{(0)}=0.649$. (b) Results for larger values of $\alpha$, at strain rates $\dot\gamma=5\times 10^{-7}$, just above their respective jamming fractions $\phi_J(\alpha)$; values of  $\phi$ are indicated in the figure.
}
\label{C-vs-y-3D}
\end{figure}

Our 3D simulations are much more time consuming and we only shear to total strains $\gamma\approx 1.4$, thus greatly reducing the number of independent samples we have to average over when computing $C(y)$.  To keep statistical accuracy reasonable, we therefore average over a thicker (as compared to 2D) layer of width $\Delta y = 0.18$ to define $n(y)$, so as to have more particles in the layer and so smaller fluctuations.
Our results for the correlations of 3D ellipsoids are shown in Fig.~\ref{C-vs-y-3D}.  While the larger $\Delta y$ means we lack the finer scale features  seen in Fig.~\ref{C-vs-y-2D} for 2D, we continue to see similar decaying oscillations, characteristic of the absence of any long range translational ordering.

Note, since we use $N=1024$ particles in both 2D and 3D, the system length for our 2D systems is $\mathcal{L}\sim 37$, while for 3D it is $\mathcal{L}\sim 11$.  Thus in 3D the oscillations in $C(y)$ have not quite decayed to zero before one reaches $y=\mathcal{L}/2$, where the periodic boundaries influence our results and give a larger $C(y)$ than would be found in a larger system.  Nevertheless our results in 3D are consistent with decaying correlations, and so the absence of any smectic ordering.

\section{Contact location distribution}
\label{contact}

In the main text we showed in Fig.~5 that the probability density per unit surface area $\mathcal{P}(\vartheta)$, for a particle to have a contact at polar angle $\vartheta$  on its surface, had a sharp peak at $\vartheta=\pi/2$, where the particle width is narrowest.  The height of this peak increases as the asphericity $\alpha$ decreases.  The results for $\mathcal{P}(\vartheta)$ vs $\vartheta$ shown in Fig.~5 were for a small strain rate $\dot\gamma$ at a  fixed packing fraction near the jamming transition for spherical particles, $\phi\approx \phi_J^{(0)}$.

\begin{figure}[h!]
\centering
\includegraphics[width=3.5in]{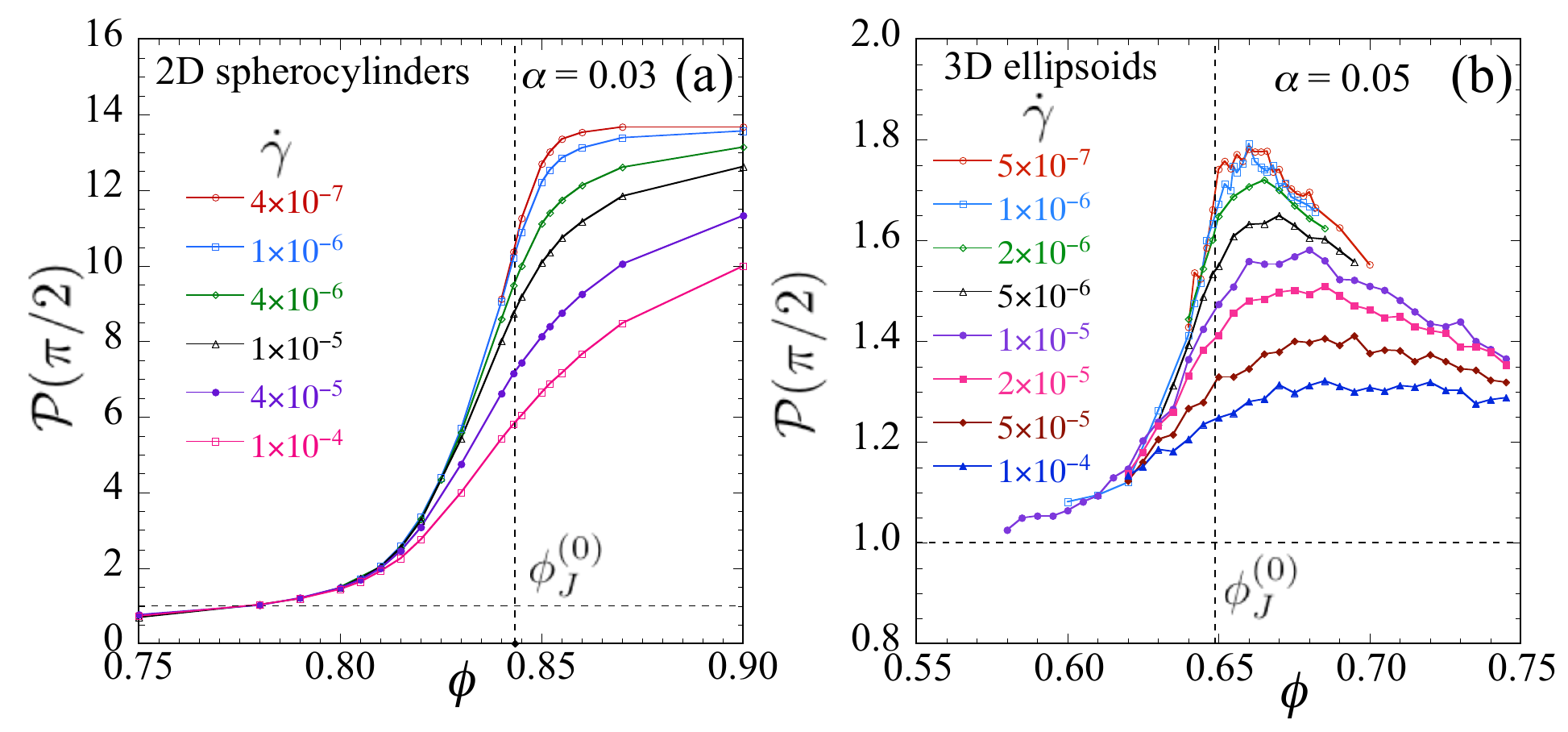}
\caption{Peak probability $\mathcal{P}(\pi/2)$ vs packing $\phi$ for  different strain rates $\dot\gamma$ for (a) 2D spherocylinders at $\alpha=0.03$ and (b) 3D ellipsoids at $\alpha=0.05$. As $\dot\gamma$ decreases, the peak value $\mathcal{P}(\pi/2)$ increases until it saturates.  Vertical dashed lines denote the jamming point of spherical particles $\phi_J^{(0)}$, while horizontal dashed lines indicate the value of unity expected for a spherical particle.
}
\label{fSM6}
\end{figure}

In Fig.~\ref{fSM6} we plot the peak height $\mathcal{P}(\pi/2)$ vs packing $\phi$ at fixed small $\alpha$, for different values of $\dot\gamma$.  
In (a) we show 2D spherocylinders at $\alpha=0.03$ and in (b) 3D ellipsoids at $\alpha=0.05$.
We see that as $\dot\gamma$ decreases, $\mathcal{P}(\pi/2)$  increases to a limiting curve, which
 rises rapidly as $\phi$ approaches $\phi_J^{(0)}$, and then stays above the spherical particle value of unity  as $\phi$ increases above the jamming transition.
 Thus the onset for the contacts to preferentially lie along the narrowest width of the particle takes place as $\phi$ passes through the jamming transition.

\section{Response to a pure shear deformation}
\label{pure}

It is interesting to compare the response of our system to a pure, rather than a simple, shear.  For simplicity we consider this for the case of our 2D spherocylinders.  
In this model the dissipative force is determined by the relative velocity of the particle with respect to an affinely deformed background host medium.  We define the local velocity $\mathbf{v}_\mathrm{host}(\mathbf{r})$ of this background host 
in terms of a strain rate tensor $\dot{\boldsymbol{\Gamma}}$, i.e., $\mathbf{v}_\mathrm{host}=\dot{\boldsymbol{\Gamma}}\cdot\mathbf{r}$.  A simple shear deformation can be decomposed into the sum of a pure shear and a uniform rotation,
\begin{equation}
\dot{\boldsymbol{\Gamma}}_\mathrm{ss}=\dot{\boldsymbol{\Gamma}}_\mathrm{ps}+\dot{\boldsymbol{\Gamma}}_\mathrm{rot}.
\end{equation}
For our coordinate system with simple shear flow in the $\mathbf{\hat x}$ direction, this becomes,
\begin{equation}
\left[
\begin{array}{cc}
0 & \dot\gamma\\
0 & 0
\end{array}
\right]
=
\left[
\begin{array}{cc}
0 & \dot\gamma/2\\
\dot\gamma/2 & 0
\end{array}
\right]
+\left[
\begin{array}{cc}
0 & \dot\gamma/2\\
-\dot\gamma/2 & 0
\end{array}
\right].
\end{equation}
The first term on the right hand side is a pure shear, with expansion along the $(1,1)$ diagonal and compression along the $(1,-1)$ diagonal, both at the rate $\dot\gamma/2$ so as to keep the area constant.  The second term is a  clockwise rotation $(-\dot\gamma/2)\mathbf{\hat z}\times \mathbf{r}$,  with angular velocity $-\dot\gamma/2$.  It is this second term which drives the continuous rotation of particles under simple shear, resulting in the finite $-\langle\omega_{zi}\rangle/\dot\gamma >0$ seen in Fig.~3 of the main text.

Under pure shear there is no such rotational drive, and particles try to relax from their initial orientation to one aligned with the expansive direction of the pure shear.
Rotating coordinates so that the expansive direction is $\mathbf{\hat x}$ and the compressive direction is $\mathbf{\hat y}$, the rotational equation of motion for pure shear becomes,
\begin{equation}
\dot\theta_i=-(\dot\gamma/2)[\Delta I_i/I_i]\sin 2\theta_i + \tau_i^\mathrm{el}/(k_dv_iI_i).
\end{equation}
For an isolated particle where $\tau_i^\mathrm{el}=0$, particles will  exponentially relax  to $\theta_i=0$ or $\pi$ with a relaxation time $t_0$ set by the total strain $\gamma_0=\dot\gamma t_0 = I_i/\Delta I_i$.  Thus, at low $\phi$ near this isolated particle limit, we expect to find near perfect nematic ordering with $S_2\approx 1$ and $\theta_2=0$.  However, as the asphericity $\alpha$ of the particles vanishes, the relaxation time needed to achieve this highly ordered state diverges as $\dot\gamma t_0=(I_i/\Delta I_i)\sim 1/\alpha$.

To investigate the response to pure shear at dense $\phi$, we have carried out numerical simulations.  A practical limitation of pure shear simulations is that, unlike for simple shear, there is a limit to the total strain $\gamma$ that can be applied to a finite numerical system before the system collapses to a narrow height of order one particle width, $L_y(\gamma)=L_y(0)\mathrm{e}^{-\gamma/2}\sim O(1)$.  To increase the total possible strain $\gamma$, we use systems of $N=1024$ particles with an initial system aspect ratio of $L_y(0)/L_x(0)=8$, and shear to a  strain $\gamma$ such that $L_y(\gamma)/L_x(\gamma)=1/8$, thus allowing a maximum  strain of $\gamma=\ln 64\approx 4.2$.  We use a strain rate $\dot\gamma=10^{-6}$, and average over four independently generated samples.

\begin{figure}[h!]
\centering
\includegraphics[width=3.5in]{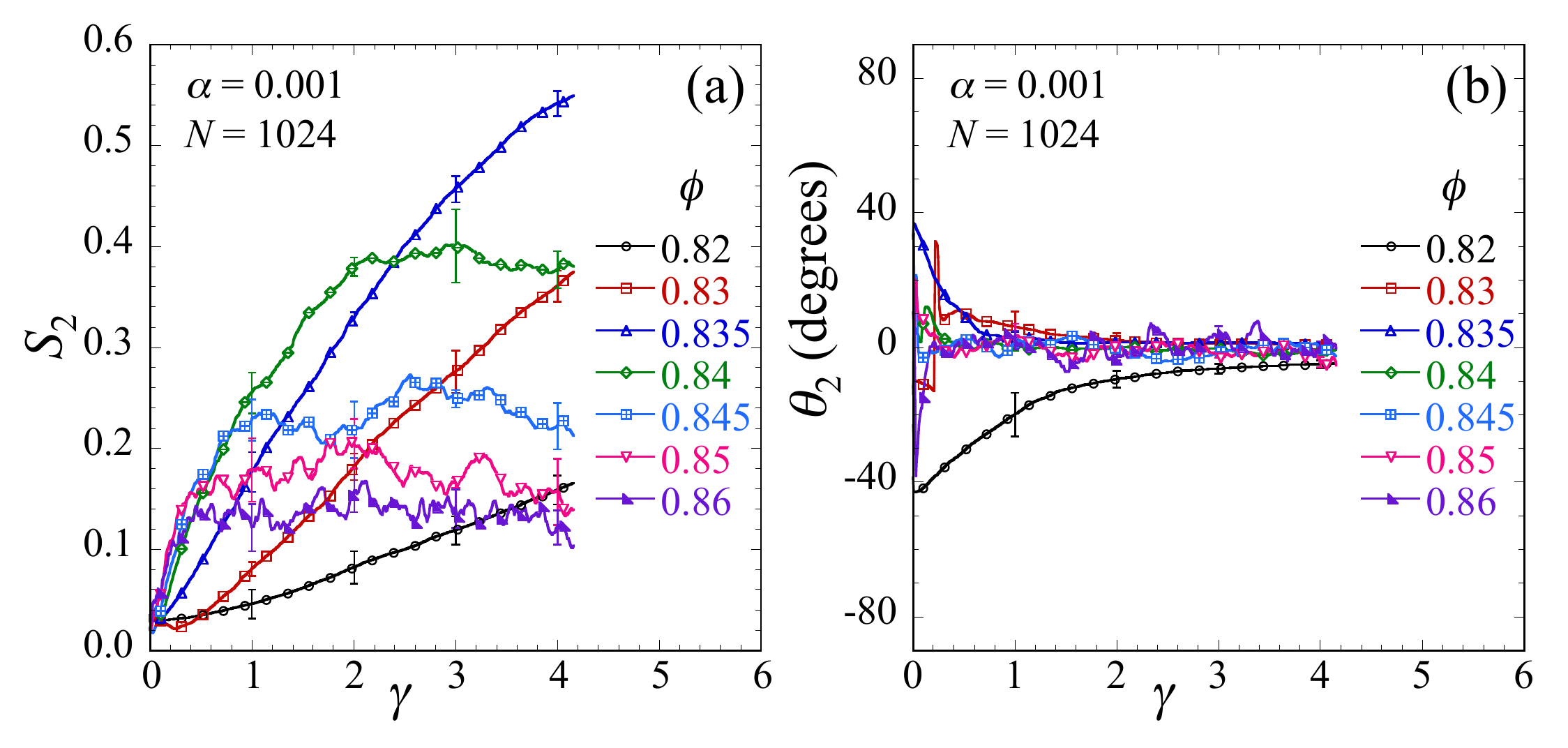}
\caption{(a) Magnitude $S_2$ and (b) direction $\theta_2$ of the nematic order parameter vs pure strain $\gamma=\dot\gamma t$, at different packing fractions $\phi$ for nearly circular particles with $\alpha=0.001$ at strain rate $\dot\gamma=10^{-6}$.  A sparse set of symbols is used to help differentiate curves of different $\phi$, with many data points existing between adjacent symbols on any curve.  Representative error bars are shown at integer values of $\gamma$.
}
\label{S2-vs-g}
\end{figure}

Here we present results for nearly circular particles at our smallest $\alpha=0.001$.  In Fig.~\ref{S2-vs-g} we plot the nematic order parameter magnitude $S_2$ and orientation $\theta_2$ vs pure shear strain $\gamma$, for several different packing fractions $\phi$.  As $\gamma$ increases, $S_2$ increases and $\theta_2$ decays from its initial random value to zero, in agreement with the expectation that particles try to relax to their preferred orientation aligned with the expansive direction $\mathbf{\hat x}$.  However
we see that we are only able to reach the desired steady state, where $S_2$ plateaus to a constant value as $\gamma$ increases, for relatively dense systems close to and above jamming, $\phi \ge 0.84$.

\begin{figure}[h!]
\centering
\includegraphics[width=3.2in]{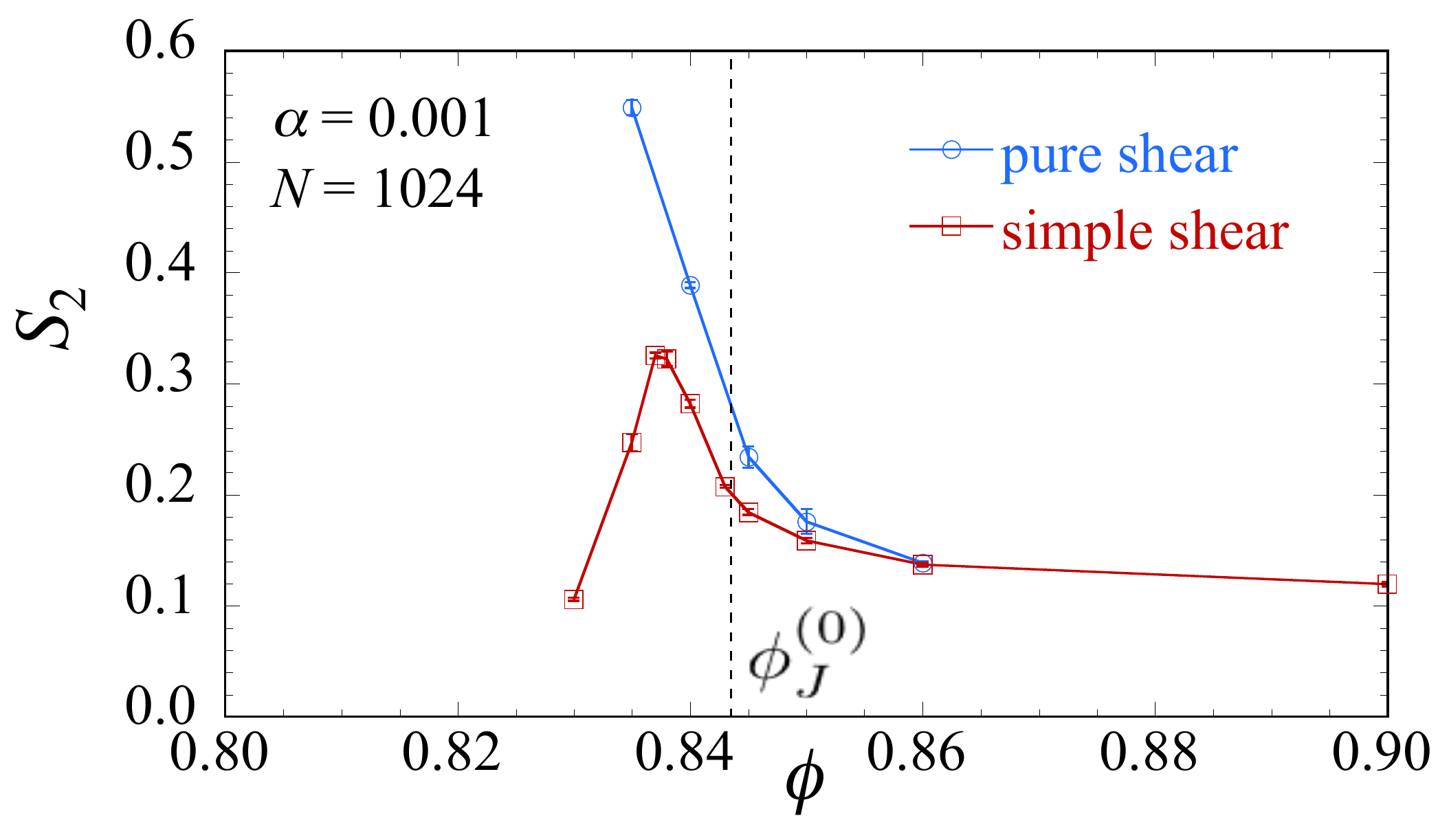}
\caption{Magnitude of the nematic order parameter $S_2$ vs packing $\phi$, comparing pure shear with simple shear, for nearly circular particles with $\alpha=0.001$.  For pure shear the strain rate is $\dot\gamma=10^{-6}$, while for simple shear $\dot\gamma=10^{-7}$.  Results represent steady state values, except for the pure shear case at $\phi=0.835$ where steady-state has not quite been reached; the value shown at this $\phi$ is therefore a lower bound on the steady state limit.
}
\label{S2-vs-phi}
\end{figure}

In Fig.~\ref{S2-vs-phi} we show the resulting steady state values of $S_2$ vs packing $\phi$, comparing results from simple shear with those from pure shear.   We see dramatically different behavior at low $\phi$.  For simple shear the nearly uniform rotation of the $\alpha=0.001$ particles results in a  small $S_2$, while for pure shear the relaxation to the expansive direction gives a large $S_2$.  As $\phi$ increases, so does the rate of particle collisions.  For pure shear the collisions and resulting excluded volume inhibit perfect alignment of particles and $S_2$ decreases. 
For simple shear the increasing collisions initially cause the rotation to slow (see Fig.~3a of the main text) and consequently $S_2$ to increase, but upon further increasing $\phi$ towards $\phi_J$ and going above, excluded volume effects similar to that in pure shear presumably inhibit alignment and cause $S_2$ to decrease, and we find that $S_2$ for both pure and simple shear become comparable and behave similarly.  The non-monotonic behavior of $S_2$ in simple shear is thus 
a consequence of the  rotational drive, present in simple shear but absent in pure shear.  However in both simple and pure shear, we find that $S_2$ at jamming remains surprisingly large, even though the particles are extremely close to circular, with the flat sides of the spherocylinders comprising only a fraction $\alpha/(\alpha+\pi/2)=6.4\times 10^{-4}$ of the particle perimeter.

\section{Determination of contacts and overlaps}
\label{deltaij}

In this section we summarize how we determine if two particles are overlapping, and if so, what is the point of contact between them.
For our 2D spherocylinders, we use the efficient algorithm of Pournin et al. \cite{Pournin} to compute the shortest distance $r_{ij}$ between the spines of two spherocylinders $i$ and $j$.  The line of length $r_{ij}$ that connects the two spines we will call the line IJ.
Whenever $r_{ij}<d_{ij}=(D_i+D_j)/2$, with $D_i$ the diameter of the endcap of spherocylinder $i$, the two spherocylinders are overlapping.  We then define the point of contact $\rr_C$, at which the elastic force acts, as the distance $[D_i/(D_i+D_j)]r_{ij}$ from the spine of spherocylinder $i$, along the line IJ.

\begin{figure}
  \centering
  \begin{tikzpicture}[scale=2.9]
  \draw[fill] (0,0) circle (0.5pt);
  \draw (0,-0.03) node[scale=1.5, below]{$\mathbf{r}_i$};
  \draw (0,0) ellipse (1 and 0.5); 
  \draw[fill] (0.8,0.4) circle (0.5pt);
  \draw (0.8,0.4+0.02) node[scale=1.5, above]{$\mathbf{r}_j$};
  \draw[rotate around={-36:(0.8,0.4)}] (0.8,0.4) ellipse (1 and 0.5); 
  \draw[dashed] (0,0) ellipse (0.655762*1 and 0.655762*0.5); 
  \draw[dashed, rotate around={-36:(0.8,0.4)}] (0.8,0.4) ellipse (0.655762*1 and 0.655762*0.5); 
  \draw[fill] (0.555670,0.174160) circle (0.5pt) {};
  \draw (0.555670+0.01,0.174160+0.01) node[scale=1.5, right]{$\mathbf{r}_C$};
  \end{tikzpicture}

  \caption{Solid lines denote two overlapping ellipsoids $i$ and $j$, with centers $\rr_i$ and $\rr_j$
    respectively; the overlap is exaggerated over what is found in the actual simulations for the sake of clarity.  
    Dashed lines show the same two ellipsoids when their axes are rescaled by a common factor $\delta_{ij}$, so that they now have a single point of contact at $\rr_C$.
}
\label{overlap}
\end{figure}
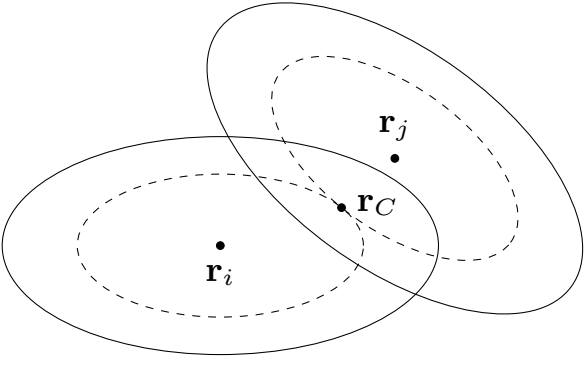

For our 3D ellipsoids, the procedure is more complicated.
As illustrated in Fig.~\ref{overlap}, for two overlapping ellipsoids $i$ and $j$ one can define a scale factor $\delta_{ij} < 1$ such
that there exists a unique point of contact $\rr_C$ between these ellipsoids when their axes are rescaled by the common factor
$\delta_{ij}$, keeping their center of mass positions fixed. 
This scale factor $\delta_{ij}$ can be computed using a method introduced by Perram and Wertheim (PW)
\cite{perram1985statistical}  which has  been  applied to the study of jammed packings of
ellipsoidal particles \cite{donev2007underconstrained, donev}. Here we  briefly summarize  this method.

For any position $\rr$, we define the scale function $\delta_i(\rr)$ such that $\rr$ will lie on the surface
of ellipsoid $i$ if its axes are rescaled by $\delta_i(\rr)$. We then introduce the contact
function $F(\rr,\lambda)$ defined for two ellipsoids $i$ and $j$, 
\begin{equation}
  F(\rr, \lambda) = \lambda \delta_i^2(\rr) + (1 - \lambda) \delta_j^2(\rr),
  \label{eq:contact_function}
\end{equation}
where $\lambda\in [0,1]$.
It has then been demonstrated \cite{perram1985statistical} that there exists
an $\rr(\lambda)$ such that
\begin{equation}
  \label{eq:nablaF}
  \bnabla F(\rr(\lambda), \lambda) = 0,
\end{equation}
where $\bnabla \equiv \partial/\partial\boldsymbol{r}$.  This  implies that
\begin{equation}
  \lambda \bnabla\delta_i^2(\rr(\lambda)) = -(1 - \lambda) \bnabla\delta_j^2(\rr(\lambda)),
  \label{eq:nabladeltai}
\end{equation}
which shows that when ellipsoids $i$ and $j$ are rescaled by factors $\delta_i(\rr(\lambda))$ and
$\delta_j(\rr(\lambda))$ respectively, the point $\rr(\lambda)$ lies on the surfaces of both ellipsoids, and the normal vectors to the surfaces at this point are parallel but pointing in opposite directions, so that the two ellipsoids are tangent at $\rr(\lambda)$.

PW further showed \cite{perram1985statistical} that $F(\rr(\lambda), \lambda)$, as a function of $\lambda\in[0,1]$, has a unique maximum at  $\lambda^*$, such that
\begin{equation}
  \begin{aligned}
       0 & =\left. \frac{d F(\rr(\lambda), \lambda)}{d\lambda}  \right|_{\lambda=\lambda^*}\\[10pt]
    & =\left. \frac{\partial F(\rr(\lambda), \lambda)}{\partial \lambda}\right|_{\lambda=\lambda^*}  \!\!+ \rr(\lambda^*) \cdot \bnabla F(\rr(\lambda^*), \lambda^*),
  \end{aligned}
\end{equation}
where the second term vanishes due to \Eq{nablaF}.  From \Eq{contact_function} we then find
$\delta_i^2(\rr(\lambda^*)) = \delta_j^2(\rr(\lambda^*))$, which means that the scale
factor is the same for both ellipsoids, and
\begin{equation}
  \label{eq:hmm}
  \delta_i^2(\rr(\lambda^*)) = \delta_j^2(\rr(\lambda^*)) = F(\rr(\lambda^*), \lambda^*).
\end{equation}
The scale factor $\delta_{ij}$ that we are seeking is thus  defined as
\begin{equation}
  \delta_{ij}^2 = \max_{ \lambda \in [0,1]} [ F(\rr(\lambda), \lambda)].
  \label{eq:scale_factor_max}
\end{equation}

With this notation, we define the point of contact between ellipsoids $i$ and
$j$ as $\rr_C = \rr(\lambda^*)$. It is thus the unique point common to
ellipsoids $i$ and $j$ when both are rescaled with a common factor $\delta_{ij}$.

To compute the scale factor $\delta_{ij}$ defined in \Eq{scale_factor_max}, we use  a
method derived from Ref.~\cite{donev}.
An ellipsoid $i$ is defined by its center of mass position $\rr_i$, the lengths of its  axes $(a_1, a_2, a_3)$, and the rotation matrix $\boldsymbol{\mathcal{Q}}_i$ that rotates the $(x, y, z)$ directions of the lab coordinate
frame onto the principal axes of the ellispoid. We then introduce the matrix,
\begin{equation}
  \B_i = \boldsymbol{\mathcal{Q}}_i\cdot
  \begin{pmatrix} a^{-2}_1 & 0 & 0 \\ 0 & a^{-2}_2 & 0 \\ 0 & 0 & a^{-2}_3 \end{pmatrix}\cdot
  \boldsymbol{\mathcal{Q}}_i^{-1},
\end{equation}
which is symmetric due to the orthogonal nature of $\boldsymbol{\mathcal{Q}}_i$, and gives an explicit definition
of the scale function $\delta_i(\rr)$,
\begin{equation}
  \delta_i^2(\rr) = (\rr - \rr_i)\cdot \B_i\cdot (\rr - \rr_i).
  \label{eq:B}
\end{equation}
\Eq{nabladeltai} then becomes
\begin{equation}
  \lambda \B_i\cdot(\rr(\lambda) - \rr_i) = - (1 - \lambda) \B_j\cdot(\rr(\lambda) - \rr_j).
  \label{eq:BiBj}
\end{equation}
After introducing \cite{perram1985statistical}
\begin{displaymath}
  \Y_{ij}(\lambda) = \lambda \B_j^{-1} + (1 - \lambda) \B_i^{-1},
\end{displaymath}
and defining $\rr_{ji}=\rr_j-\rr_i$,
\Eq{BiBj} gives expressions for the distances between the contact point $\rr(\lambda)$ and
the centers of the ellipsoids, 
\begin{equation}
  \begin{aligned}
    \rr(\lambda) - \rr_i &= (1 - \lambda) \B_i^{-1}\cdot\Y^{-1}_{ij}(\lambda)\cdot\rji,\\
    \rr(\lambda) - \rr_j &= -\lambda \B_j^{-1}\cdot\Y^{-1}_{ij}(\lambda)\cdot\rji.
  \end{aligned}
  \label{eq:rlambda-r}
\end{equation}
As discussed above, the unique contact point $\rr_C$ for equal scale factors, $\delta_{ij}=\delta_i(\rr_C)=\delta_j(\rr_C)$, 
is found by maximizing the
contact function $F(\rr(\lambda),\lambda)$ with respect to $\lambda$.  Using the above results and \Eq{B} in
\Eq{contact_function} thus gives,
\begin{eqnarray}
\nonumber
  F(\rr(\lambda), \lambda) & = &
  \lambda (1 - \lambda) \rji\cdot \Y^{-1}_{ij}(\lambda)\cdot \rji \qquad\qquad
  \\[10pt]
  & = & \frac{\lambda(1 - \lambda)\rji\cdot \mathrm{adj}[\Y_{ij}(\lambda)]
    \cdot\rji}{\mathrm{det}[\Y_{ij}(\lambda)]}\\[8pt]\nonumber
  &\equiv& \frac{p_{ij}(\lambda)}{q_{ij}(\lambda)},
\end{eqnarray}
where $\mathrm{adj}[\ldots]$ denotes the adjugate matrix (whose element $(\alpha,\beta)$ is
equal to the determinant of the $2\times2$ sub-matrix obtained after eliminating  row $\beta$
and column $\alpha$ from the original $3\times 3$ matrix), and $\mathrm{det}[\ldots]$ denotes the
determinant.  The functions 
$p_{ij}(\lambda)$ and $q_{ij}(\lambda)$ are polynomials in $\lambda$ of degree 4 and 3
respectively.

The desired parameter $\lambda^*$, at which $F(\rr(\lambda), \lambda)$ is  maximized, is then the unique root 
in the interval $[0, 1]$ of the 6th degree polynomial 
\begin{equation}
h_{ij}(\lambda) = p'_{ij}(\lambda)q_{ij}(\lambda) - p_{ij}(\lambda)q'_{ij}(\lambda),
\end{equation}
i.e., $h_{ij}(\lambda^*)=0$, where primes above denote derivatives with respect to $\lambda$.  

Finally, to determine ellipsoid elastic interactions, we investigate all pairs of ellipsoids whose center of mass separation $|\rr_i-\rr_j|$ is small enough that the ellipsoids might be overlapping.  We then apply the above procedure to determine $\delta_{ij}$.  If the resulting $\delta_{ij}>1$, then the pair of ellipsoids are in fact not overlapping and so have no interaction.  If $\delta_{ij}\le 1$, then the ellipsoids overlap and the point of contact is taken as $\rr_C$.


\begin{thebibliography}{99}
\bibitem{comment}For spherocylinders it is  ambiguous how to define the location of a contact when two flat sides are touching.  Here we  take such a contact to have unit strength but distribute its location evenly over the segments of the flat surfaces that are in contact.   For small $\alpha$ our results are insensitive to this as compared to other choices.
\bibitem{contacts}For 2D spherocylinders we find that near $\phi_J$ the integrated probability for a contact to lie anywhere along the flat sides stays constant as $\alpha\to 0$, $\dot\gamma\to 0$.
\bibitem{poly}Preliminary results for 2D spherocylinders suggest that our conclusions remain unchanged even with polydispersity in $\alpha$.
\end{thebibliography}
\end{document}